 \documentclass[]{acmart}

\usepackage{longtable}
\AtBeginDocument{%
  \providecommand\BibTeX{{%
    \normalfont B\kern-0.5em{\scshape i\kern-0.25em b}\kern-0.8em\TeX}}}

\copyrightyear{2024}
\acmYear{2024}
\setcopyright{acmlicensed}
\acmConference[IDC '24]{Interaction Design and Children}{June 17--20, 2024}{Delft, Netherlands}
\acmBooktitle{Interaction Design and Children (IDC '24), June 17--20, 2024, Delft, Netherlands}
\acmDOI{10.1145/3628516.3655752}
\acmISBN{979-8-4007-0442-0/24/06}

\begin{document}

\title[Youth as Peer Auditors]{Youth as Peer Auditors: Engaging Teenagers with Algorithm Auditing of Machine Learning Applications}

\author{Luis Morales-Navarro}
\email{luismn@upenn.edu}
\orcid{0000-0002-8777-2374}
\affiliation{%
  \institution{University of Pennsylvania}
  \city{Philadelphia}
  \state{PA}
  \country{United States}
}

\author{Yasmin B. Kafai}
\email{kafai@upenn.edu}
\orcid{0000-0003-4018-0491}
\affiliation{%
  \institution{University of Pennsylvania}
  \city{Philadelphia}
  \state{PA}
  \country{United States}
}

\author{Vedya Konda}
\email{vedyask@upenn.edu}
\orcid{0009-0006-5776-5532}
\affiliation{%
  \institution{University of Pennsylvania}
  \city{Philadelphia}
  \state{PA}
  \country{United States}
}

\author{Danaë Metaxa}
\email{metaxa@seas.upenn.edu}
\orcid{0000-0001-9359-6090}
\affiliation{%
  \institution{University of Pennsylvania}
  \city{Philadelphia}
  \state{PA}
  \country{United States}
}

\renewcommand{\shortauthors}{Morales-Navarro, Kafai, Konda, \& Metaxa}

\begin{abstract}
  As artificial intelligence/machine learning (AI/ML) applications become more pervasive in youth lives, supporting them to interact, design, and evaluate applications is crucial. This paper positions youth as auditors of their peers' ML-powered applications to better understand algorithmic systems' opaque inner workings and external impacts. In a two-week workshop, 13 youth (ages 14-15) designed and audited ML-powered applications. We analyzed pre/post clinical interviews in which youth were presented with auditing tasks. The analyses show that after the workshop all youth identified algorithmic biases and inferred dataset and model design issues. Youth also discussed algorithmic justice issues and ML model improvements. Furthermore, youth reflected that auditing provided them new perspectives on model functionality and ideas to improve their own models. This work contributes (1) a conceptualization of algorithm auditing for youth; and (2) empirical evidence of the potential benefits of auditing. We discuss potential uses of algorithm auditing in learning and child-computer interaction research.
\end{abstract}





\maketitle

\section{Introduction}
Today, children and youth interact with artificial intelligence/machine learning (AI/ML)-powered applications and algorithmic systems when they socialize with friends, go to school, play games, listen to music, do homework, order food, or watch videos. Given the increasing prevalence of AI/ML in their lives, it is crucial to provide young people with the necessary support to engage with, create, and evaluate AI/ML applications. As such, child-computer interaction (CCI) research on AI/ML literacy has received increasing attention \cite{long2020ai, vartiainen2021machine, irgens2022characterizing}. An obstacle in supporting young people in understanding and engaging with AI/ML ideas is the lack of transparency in ML models. Furthermore, existing research gives little attention to critical issues of computational empowerment \cite{dindler2020computational} such as supporting youth in thinking about the limitations and implications of AI/ML technologies \cite{van2023emerging} or in considering algorithmic justice; that is, how algorithmic systems may be ineffective, even perpetuate harm, and disproportionately impact vulnerable people \cite{birhane2021algorithmic}.

In human-computer interaction (HCI) and algorithmic justice research, an effective strategy for investigating and understanding the opaque inner workings and implications of AI/ML systems is algorithm auditing. Algorithm auditing is a method introduced about a decade ago that involves ``repeatedly querying an algorithm and observing its output in order to draw conclusions about the algorithm’s opaque inner workings and possible external impact'' \cite[p. 272]{metaxa2021auditing}. Most of these audits are conducted with the goal of identifying problematic system behaviors in AI/ML-powered systems \cite{bandy2021problematic}. But to date, most research on algorithm auditing has focused on experts, with a few recent studies on how non-expert adults engage with the method.

In this paper, we investigate the role of youth as auditors of ML-powered applications by building on CCI’s rich tradition of exploring the various roles young people can have in contributing to the design of computing applications \cite{fails2013methods, lehnert2022child}. We conducted a two-week workshop with 13 youth (ages 14-15) in which they designed and audited each other's ML applications. We analyzed pre and post clinical interviews in which youth were presented with auditing tasks to address the following research questions: 
\begin{itemize}
  \item How did youth’s identification of potential algorithmic biases and harm change from pre to post? 
  \item How did youth’s inferences about data and model design change from pre to post?
\end{itemize}
In the post interviews, we also analyzed students reflections about auditing activities during the workshop to address:
\begin{itemize}
  \item What benefits did youth find in auditing applications and having their applications audited?  
\end{itemize}
Our analysis revealed that in post, all participants identified potential algorithmic biases and made inferences about dataset and model design issues. In post, more youth talked about algorithmic justice and next steps to further improve ML models. Furthermore, participants reflected that auditing provided them with new perspectives on model functionality and ideas to improve their own models. This paper contributes (a) a conceptualization of algorithm auditing for youth, adapting methods used with adults in algorithmic justice research by grounding them in the rich history of child-computer interaction research; and (b) an empirical clinical pre/post interview study in which youth completed auditing tasks. We discuss future directions for incorporating algorithm auditing in learning activities and CCI research as a promising practice to promote computational empowerment \cite{dindler2020computational}.

\section{Background}

Child-computer interaction (CCI) has been concerned with the different roles that young people can play in the design of computing applications since its early days \cite{fails2013methods, lehnert2022child}. This has led to the development of various rich methods to involve children in design processes as informants \cite{dowthwaite2020s, scaife1999kids}, design partners \cite{druin2002role, yip2023co}, testers \cite{lamichhane2023children, solomon2020history}, and designers \cite{holbert2020designing, kafai1991learning} (for a detailed review of the theories and methods driving the participation of children and youth in the design process, see \cite{fails2013methods}). In the following sections, we delve into how CCI has addressed the role of children as testers and evaluators at large and in the context of AI/ML. Following, we address how auditing differs from other testing and evaluation methods, review current research on non-expert auditing, and work on youth’s perspectives on algorithmic justice to propose positioning youth as auditors of their peers’ applications. 

\subsection{Youth as testers and evaluators}

Research on testing and evaluation can be grouped into two broad categories: (1) when children test and evaluate applications created by experts; and (2) when children test and evaluate child-designed applications. Druin \cite{druin2002role} defines the role of testers as users who also help identify ``design and usability issues for revision of prototypes.'' Engaging children and youth as testers in the design process of technologies created by experts can be traced back to Solomon and Papert’s work on LOGO in the mid-1960s, when they conducted year-long iterative test sessions with children to refine the design of the programming language \cite{solomon2020history, druin2002role}. Since then, children have been involved as testers in the design of Smalltalk \cite{goldberg1979educational, kay1996early}, Scratch \cite{maloney2008programming}, and most child-facing applications. Traditionally, when experts lead the design, it is adult researchers that interpret how children tested the technologies and synthesize the findings of testing sessions \cite{fails2013methods}. More recent work in CCI has engaged children in testing tangible interfaces for learning mathematics \cite{zito2021leveraging}, as well as the development of instruments to better understand children’s engagement when testing applications \cite{dietz2020giggle}. In terms of children as testers of AI/ML systems, some work has been conducted with children testing a machine translation application \cite{lamichhane2023children} as well as an application for learning about reinforcement learning \cite{dietz2022artonomous}.  

A second strand of research on children as testers and evaluators involves child-designed applications. This work can also be traced back to LOGO, in particular Kafai and Harel’s late 1980s research on children as designers of software \cite{kafai1991learning}. Positioning children as designers of instructional software for learning mathematics \cite{harel1991children} and video games \cite{kafai2012minds}, they created environments in which children could test their software with their peers. Additionally, older peers could take on the role of "consultants" or outside evaluators who examined peer-created software and, by "playing doctor," assisted in identifying and diagnosing problems \cite{kafai1991consult}. While being consultants, children benefited from cognitive distance and were able to provide designers with new perspectives, refining their understanding of problem behaviors. Later work \cite{kafai1998children} looked at the differences between designer-led usability testing and external evaluation, highlighting that when peers play the role of testers in designer-led tasks, it is the designers that benefit from gaining insights on how to improve their own projects to meet the needs of their users. On the other hand, external evaluation of software provided opportunities for evaluators to “apply the insights gained from their own design process” \cite[p. 128]{kafai1998children}. These research studies highlight how testing and evaluating applications can also support youth in their learning of computing. Since then having children design artifacts that can be tested by their peers has become a common activity in many CCI projects.  

Several CCI studies have mentioned the importance of engaging youth in testing AI/ML models, but they often lack detailed findings on how young people evaluate models and what they can gain from the testing process \cite{voulgari2021learn, kaspersen2022high, hitron2019can, burhans2017arty}. This is also the case in the AI/ML education literature, where training models has received the most attention \cite{morales2024not}. The few studies that have investigated how youth test their own models show promising results. Several studies argue that when youth test models, they build hypotheses and explanations for model behaviors and also come up with new ideas for how model performance could be improved \cite{vartiainen2020learning, vartiainen2021machine, druga2021children}. Recent work shows that testing can support young people in identifying issues related to data diversity, class imbalance, and data quality \cite{tseng2023co, tseng2023collaborative}. Yet testing is not always systematic and in-depth. Sometimes young people, after identifying cases in which models do not perform as expected, instead of trying to fix the models, change their testing practices \cite{zimmermann2019youth}. Other studies have shown that youth rarely test their models, only doing it when prompted by researchers, or that sometimes they think that by simply testing they can improve model performance without making changes to training datasets, model parameters, or retraining \cite{zimmermann2020youth}. A couple of studies also engage youth as testers of their peers' models in designer-led testing activities \cite{dwivedi2021exploring, morales2024not}, that is, when the designers of the applications guide their peers in the testing process. Notably, none of these studies involved youth as external evaluators that evaluate models from the outside in.

What we learn from these previous studies is that there is already a rich tradition in CCI research of engaging children and youth as testers and evaluators, from traditional software to machine learning applications. In introducing youth as auditors, we are adding a new ``role'' to the repertoire that is distinct from previously examined roles. Here we describe how auditing is different from other forms of testing and evaluation, review existing research that involves non-experts in algorithm auditing, and research on algorithmic justice and youth that inform our approach to positioning youth as peer auditors. 

\subsection{Youth as auditors of AI/ML applications}

To begin, we note that auditing differs from traditional testing and evaluation in several ways \cite{metaxa2021auditing}. In algorithm auditing, traditionally, the emphasis is on the system itself rather than how users react to or interact with it, though recent work is beginning to include users as part of the system being audited \cite{lam2023sociotechnical}. Unlike other forms of testing, auditing is systematic, with the intention of drawing conclusions at the level of the system rather than about individual test cases. Finally, audits are generally external evaluations done by independent third parties from the outside-in, based on externally-measured system behaviors. 

Traditionally, teams of expert auditors conduct audits using methods such as scrapping, automatically collecting and analyzing data from online sources, or “sock-puppets,” in which researchers collect data by imitating user behaviors \cite{bandy2021problematic}. For example, an expert audit by Metaxa and colleagues \cite{metaxa2021image} investigated gender and racial representation disparities in Google Images by scrapping and analyzing image search results. They found evidence of under-representation of women and people of color in queries of common job occupations in search relative to the U.S. workforce. Some other audits involve non-experts through crowdsourcing, collecting data in distributed and centralized ways. Here, the involvement of non-experts, for example, could include asking users to install a browser extension that automatically queries a system and logs the resulting data \cite{robertson2018auditing}. 

Recently, HCI researchers have started investigating how non-experts engage with algorithm auditing by involving them beyond crowdsourcing data. Studies have looked at how users engage in emergent, everyday auditing practices, without the participation of experts, on social media platforms \cite{shen2021everyday}. Other work has investigated approaching auditing from a socio-technical perspective by auditing both system and non-expert auditor practices \cite{lam2023sociotechnical, lam2022end}. These studies highlight that non-expert audits can uncover problematic algorithmic behaviors that experts may not be able to find \cite{lam2022end, shen2021everyday}. Notably, all of these studies have been conducted with adults. Another important finding has been the collaborative nature of non-expert auditing which often involves sharing problematic findings with others via social media—a practice that might connect well with youth engaged in similar tasks.

More closely related to our work, DeVrio and colleagues \cite{devos2022toward} have investigated how non-expert adults involved in auditing tasks make sense of potentially harmful behaviors in algorithmic systems. For instance, they had users conduct Google image searches during think-aloud interviews in which participants were tasked with looking for specific images using keywords that may show potential harmful biases. Following, they asked participants to search for other keywords that may also generate problematic results. The study showed that users' findings and interpretations are based on their prior experiences and exposure to societal biases. Furthermore, users came up with ideas to reduce harmful biases, including increasing representation diversity in the content and in the order in which results are displayed. 

\subsubsection{\textbf{Youth’s perspectives towards algorithmic justice}}
While, to our knowledge, no studies have investigated the role of youth as auditors, a handful of studies have researched youth’s perspectives towards algorithmic justice and potential harmful biases. Researchers have engaged youth in discussions in relation to high-stakes policing surveillance technologies \cite{vakil2022youth} and hypothetical robot interactions \cite{charisi2021exploring}. For instance, Coenraad \cite{coenraad2022s} and Salac et al. \cite{salac2023scaffolding, Salac2023Funds} investigated youth’s perceptions of algorithmic fairness. They found that youth ``demonstrated an awareness of visible negative impacts of technology and provided examples of this bias within their lives” \cite{coenraad2022s} but did not have the words to discuss bias or how ``invisible bias" emerged. After introducing examples of threats to equity, youth were able to discuss visible and invisible issues of equity. Salac and colleagues \cite{salac2023scaffolding} presented children and youth with scenarios of algorithmic unfairness to prompt their understandings of how the systems worked. The scenarios included bias towards female nurses in image search, a voice assistant not understanding a student with an accent, and a case of inequitable access to school supplies. They explain that children used human and technical lenses to make sense of the issues they were presented with and, at the same time, brought up their own identities and lived experiences in discussing the scenarios. Teenagers examined potential sources of bias and considered the effects these could have in different contexts and on individual people as well as communities. 

Solyst and colleagues \cite{solyst2023would} have also investigated youth’s perspectives towards fairness, finding that youth have a desire for agency to participate in the design of technology and define how applications should work. In a different study, they engaged youth in activities to identify algorithmic biases and propose ways to address these \cite{solyst2023potential}. Here participants interacted with examples of image search on Google and image generation in DALL-E, finding that youth identified various types of biases and different potential harms that these could cause. Furthermore, in computing education research, audits have been discussed for their potential as productive opportunities for critical inquiry in which learners investigate the limitations and implications of computing applications \cite{morales2023conceptualizing}. For example, inspired by algorithm auditing research, Walker and colleagues \cite{walker2022liberatory} adapted Buolamwini’s \cite{buolamwini2022facing} ideas about “evocative audits” into a learning activity in which young African American students used art to reflect on the harm that algorithms may inflict on their communities.

\subsubsection{\textbf{A new role for youth}}

The previous research findings on algorithm auditing with non-experts and current work on algorithmic justice and youth provide us with a promising foundation to conceptualize the role of youth as auditors in the tradition of CCI research. Here, it is possible to imagine different ways in which youth could be positioned as auditors of applications. For instance, building on expert audit research on sock puppet auditing \cite{bandy2021problematic}, youth could be guided to learn about auditing by creating fictional personas to collect data and evaluate how systems behave differently depending on who uses them. Building on non-expert auditing work on emergent audits in social media \cite{shen2021everyday}, CCI researchers could investigate how youth audit popular applications (such as TikTok) both in “the wild” and in auditing workshops. Similarly to DeVrio, Solyst, and Salac’s work \cite{devos2022toward, salac2023scaffolding, solyst2023potential}, CCI researchers could design and co-design tools and learning activities to engage youth in auditing the technologies they use in their everyday lives. Finally, building on AI/ML testing activities \cite{morales2024not, tseng2023co, dwivedi2021exploring} youth could audit each other's applications. We further discuss this approach in the next subsection.

\subsubsection{\textbf{Youth as peer auditors}}
\begin{figure*}[ht]
  \centering
  \includegraphics[scale=0.7, width=\linewidth]{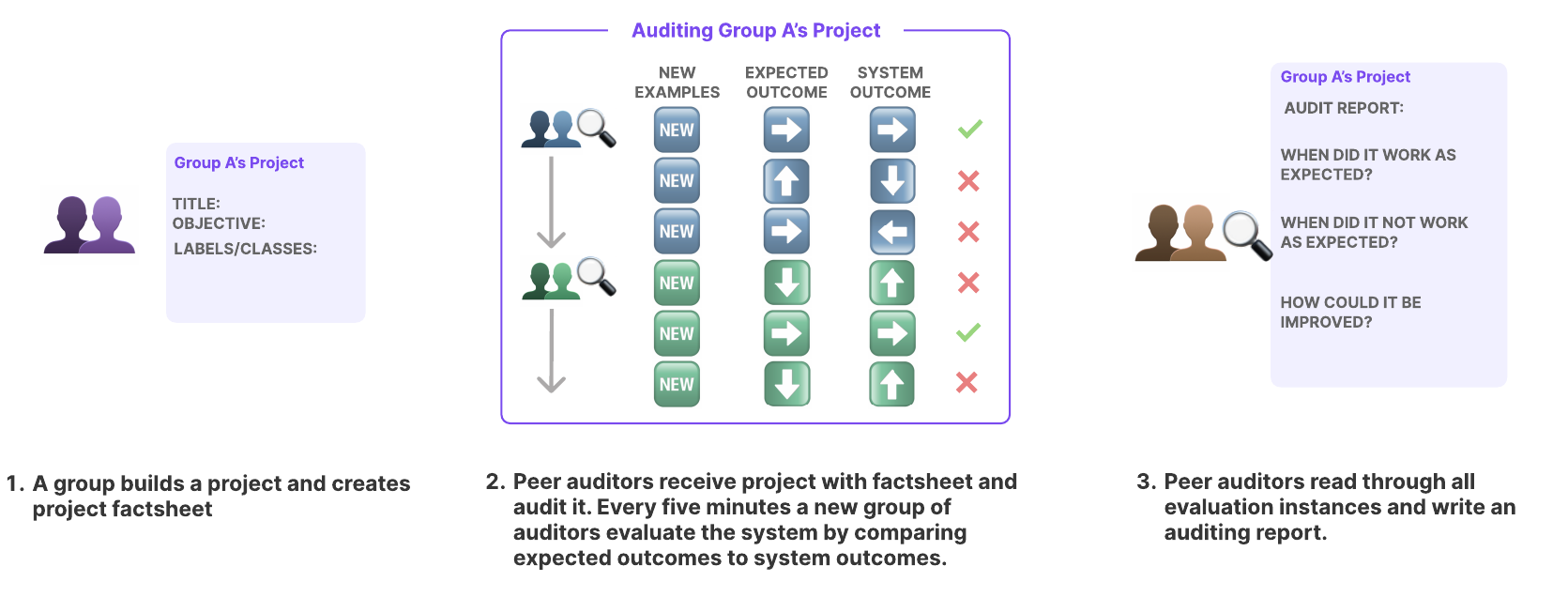}
  \caption{The process of peer auditing involves: (1) youth creating factsheets for the projects they designed, (2) exchanging projects with peers for auditing, which involves iteratively querying the system and comparing auditor-expected classification outputs to system classification outputs, and (3) writing audit reports.}
  \Description{A group builds a project and creates a project factsheet. Peer auditors receive a project with a factsheet and audit it. Every five minutes, a new group of auditors evaluate the system by comparing expected outcomes to system outcomes. Peer auditors read through all evaluation instances and write an audit report. }
  \label{fig:peer}
\end{figure*}

In this paper, we examine positioning youth as peer auditors of AI/ML applications. As peer auditors, youth can audit applications designed by their peers by collaboratively and iteratively querying the systems to evaluate their behaviors against expected behaviors. Like the consultants in the LOGO studies, youth can "play doctor" and assist in identifying and diagnosing problems \cite{kafai1991consult}. Playing the role of an auditor may have similar benefits to those already identified when youth test their own applications, including identifying issues related to data diversity, class imbalance, and data quality, building hypotheses and explanations for model behaviors, and coming up with new ideas for how model performance could be improved \cite{vartiainen2020learning, van2023emerging, morales2024not, druga2021children, tseng2023co}. In the case of classifiers, the context of this study, peer auditing involves iteratively querying the system, comparing auditor-expected classification outputs to system classification outputs, and analyzing the results to make inferences about system behavior (see Fig. \ref{fig:peer}).

We conducted a workshop in which youth first designed applications that used classifiers. Following, they wrote project factsheets that specified the objectives of the projects and their labels/classes. Then, they exchanged projects with their peers. Peer auditors iteratively queried the systems and documented each query. To ensure a wide variety of queries, every five minutes the auditors of the projects changed, with youth rotating through all projects but their own. Finally, auditors analyzed the data gathered and wrote a report in which they were asked to describe when the applications worked as expected, when they observed unexpected behaviors, and possible next steps to improve system behaviors.  

To evaluate the potential benefit of peer auditing activities, before and after the workshop, we conducted a pre/post clinical interview study in which participants were presented with auditing tasks and asked to explain what they were thinking as they completed them \cite{disessa2007interactional}. 

\section{Methods}
\subsection{Participants}
We held a two-week in-person workshop at a science center in the Northeastern United States with fifteen youth (ages fourteen to fifteen) who had shown interest in STEM by taking part in an after-school program meant to increase participation for historically underrepresented communities. Thirteen obtained guardian consent and assented to participate in research. Participants were already acquainted with one another, having participated in the science center program for at least a year. Out of the participants, six identified as female and seven as male. Of the participants, seven identified as Black, five identified as White, three as Latinx, two as Asian, and one as Native American, with five choosing more than one category. Eleven participants had taken computing classes at school or attended out-of-school CS workshops. None had taken workshops or courses on AI/ML (see Table \ref{table:demo}). Science center staff sent out paper handouts and emails inviting youth to take part in the study. Before the study began, guardians completed consent forms that included a brief explanation of the research, and youth gave their assent to participate. The institutional review board of the university approved the study protocol. All names mentioned in the paper are pseudonyms. 

\begin{table*}[ht]
\caption{Self-reported demographic information.}
\begin{tabular}{lllll}
\hline
\textbf{Pseudonym} & \textbf{Age} & \textbf{Gender} & \textbf{Race \& Ethnicity} & \textbf{Previous CS experience} \\ \hline
Kayla              & 14           & Female          & Black                      & Yes                             \\ \hline
Lou                & 15           & Female          & Black                      & No                              \\ \hline
Jerome             & 15           & Male            & Native American \& Black   & Yes                             \\ \hline
Bryan              & 15           & Male            & Asian \& White             & Yes                             \\ \hline
Jackie Star        & 15           & Female          & White                      & Yes                             \\ \hline
Fatimah            & 14           & Female          & Black                      & Yes                             \\ \hline
Andrés             & 14           & Male            & Latinx                     & Yes                             \\ \hline
Richard            & 14           & Male            & White                      & Yes                             \\ \hline
Iván               & 14           & Male            & Latinx \& White            & No                              \\ \hline
Emily              & 14           & Female          & Black                      & Yes                             \\ \hline
Luke               & 15           & Male            & Black \& Latinx            & Yes                             \\ \hline
Stephanie          & 15           & Female          & Black \& White             & Yes                             \\ \hline
Walter             & 15           & Male            & Asian                      & Yes                             \\ \hline
\end{tabular}
\label{table:demo}
\end{table*}

\subsection{Workshop activities}
During the workshop, participants learned about ML in the context of designing, testing, and auditing physical computing (e-textiles in particular) applications. Each workshop session had a duration of 3.5 hours, which included a 30-minute community-building activity and a 15-minute snack break. In the first week of the workshop, youth participated in structured activities to learn about machine learning classifiers, e-textiles, and how to create projects that incorporate ML and physical computing. The physical computing activities provided practical experience for youth to learn how to program the micro:bit microcontroller, use sensors and actuators, construct circuits, and sew with conductive thread. Afterwards, youth participated in hands-on activities to learn about AI/ML, different types of models, the ML pipeline \cite{fiebrink2019machine}, and data design practices \cite{tseng2023co} for training and testing image, audio, and pose classifiers created using ml5.js (a beginner-friendly machine learning javascript library), as well as Teachable Machine and a similar application for training and testing models with sensor data. They then used Bluetooth to send the classifiers' outputs to the micro:bit. 

Auditing played an important part in the work workshop durin both weeks. On the fourth day of the first week, participants were introduced to algorithm auditing and participated in an auditing activity for an image classifier they had designed. For this activity, youth in pairs first prepared a factsheet describing the expected behavior of their project and then handed it over to their peers for auditing. After receiving a project, youth proceeded to evaluate their peers’ classifiers, and every five minutes, they exchanged projects to have a wide range of auditors evaluate the projects from the outside in. While auditing, they kept track of individual testing instances on a table. Finally, youth wrote an audit report for the designers of the projects in which they synthesized their findings and made recommendations on how to improve the model. During the second week, we had another auditing session in which, following the same format as in week one, participants audited each other's projects and created audit reports. 

\begin{figure*}[ht]
  \centering
  \includegraphics[scale=0.7, width=\linewidth]{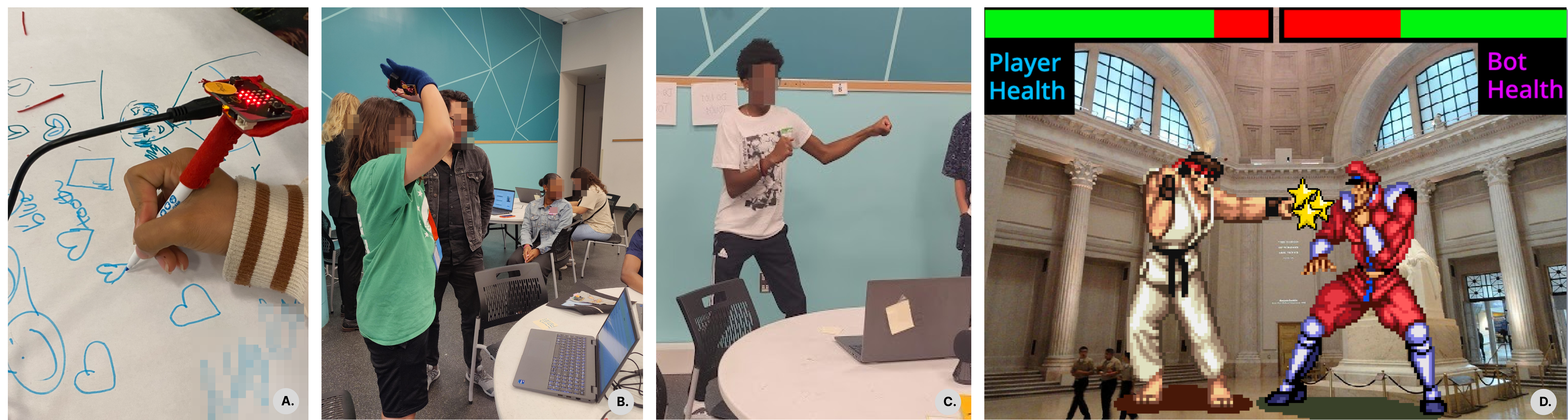}
  \caption{ Auditing youth's final projects. (A) A participant audits Jackie Star and Emily's drawing game. (B) A participant audits Andrés sports game. (C \& D) A participant audits Iván and Walter's fighting game.}
  \Description{Young person draws a heart to audit the performance of the drawing game. A young person audits the performance of a basketball game with a slam dunk. A young person audits fighting game with superman punch.}
  \label{fig:prj}
\end{figure*} 

To illustrate the activities of the workshop, we describe some of the final projects and peer auditors’ key findings. Jackie Star and Emily created a drawing game (Fig. \ref{fig:prj} A). The game involved players trying to match drawings displayed on the screen of a micro:bit attached to a pen. The project used an image classifier to classify drawings made by users; if the user drew the right shape, the micro:bit played celebratory music and prompted the player with another shape. Among other issues, auditors identified that the project did not work well with “curvy squares that don’t have super sharp angles.” Andrés created a sports game that detected different basketball moves. He attached the micro:bit to a glove and used data from its accelerometer to train a move classifier (Fig. \ref{fig:prj} B). Auditors found that the project constantly misclassified moves when users were six feet or taller. Iván and Walter created a fighting game (Fig. \ref{fig:prj} C \& D) that imitated Mortal Combat, where users controlled their players by kicking, doing uppercut punches, or superman punches. The game recognized and classified poses. Auditors noted that the game only worked well when played against plain white walls and when only one person was in the frame.

\subsection{Interview design and data collection }

We conducted pre interviews a week before the workshop and post interviews on the last day of the workshop. The interviews consisted of two pre-determined auditing tasks to evaluate image classifiers and a text-to-image generative model, accompanied by prompts that were specifically designed to elicit students' ideas (e.g., about bias, data and model design, and justice) in an open-ended manner. The interviews were conducted in pairs and deliberately structured to resemble conceptual change research interviews \cite{disessa2007interactional, lee2017rubric, sherin2012some}. Each interview had a mean duration of 23 minutes, with individual interviews ranging from 12 minutes to 31 minutes. We recorded audio and participant screens during the interviews. 

\begin{figure*}[ht]
  \centering
  \includegraphics[scale=0.7, width=\linewidth]{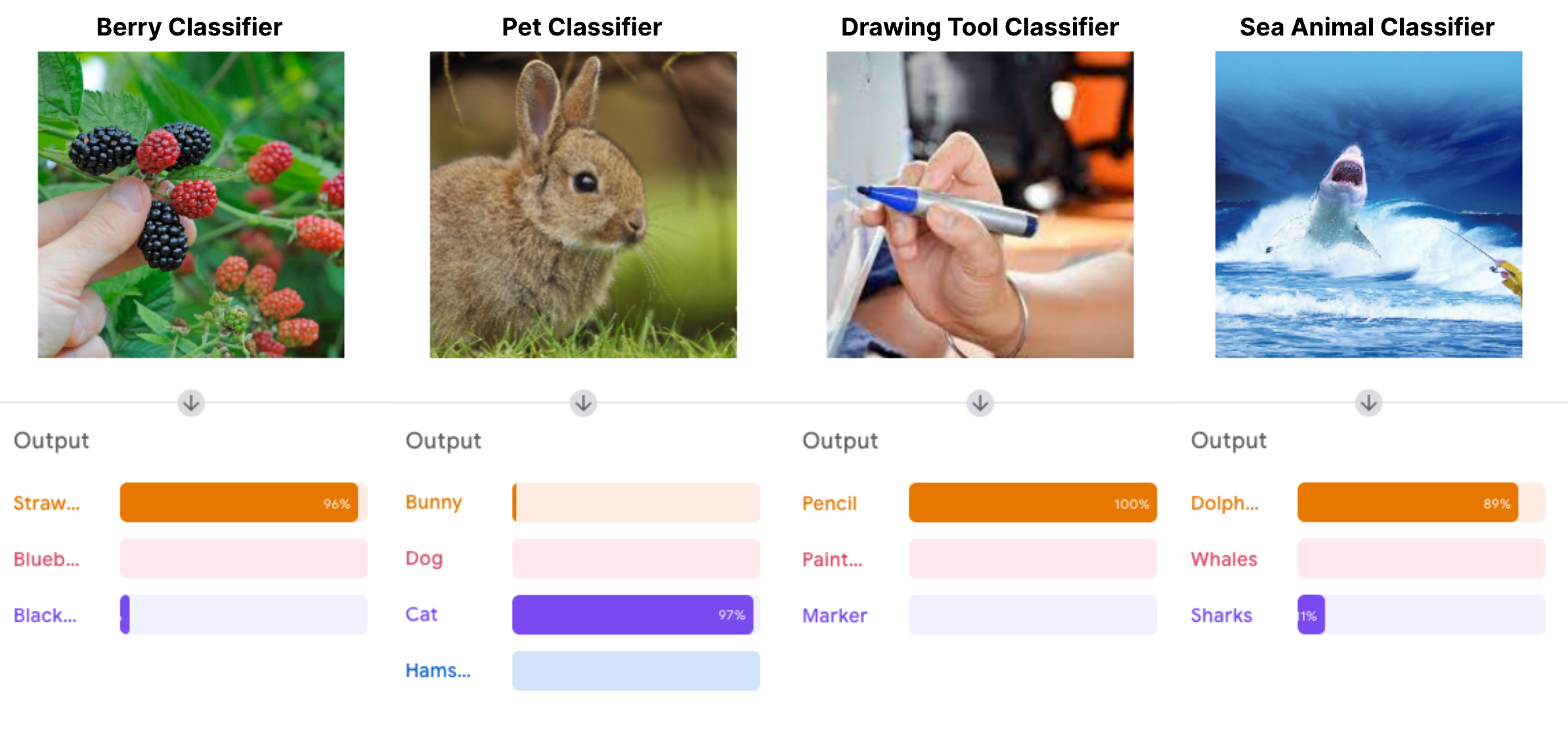}
  \caption{Faulty classifiers used in pre/post auditing tasks included a berry classifier, a pet classifier, a drawing tool classifier, and a sea animal classifier. The figure shows cases in which the classifiers did not work as expected.}
  \Description{A blackberry is misclassified as strawberry, a brown bunny is misclassified as a cat, a marker is misclassified as a pencil, and a shark is misclassified as a dolphin.}
  \label{fig:classifiers}
\end{figure*} 

In each interview, we presented youth with two image classifiers that were intentionally faulty. This task was adapted from prior research on ethics in AI/ML education in which youth were presented with faulty cat and dog classifiers and their datasets \cite{ali2019constructionism}. We prepared classifiers with inter- and intra-class imbalances (e.g., in the berry classifier, the training data included pictures of strawberries in all shapes, sizes, and colors, while blueberries and blackberries were limited to a few very similar pictures), spurious relationships (e.g., in the drawing tools classifier, all pictures of pencils in the training data included human hands and none of the pictures of other tools included hands; pictures of markers or paint brushes with hands were misclassified), and overfitting issues (e.g., in the pet classifier, all bunnies in the training data were white, as such, bunnies of other colors were misclassified). During the interviews, we first asked participants to interact with the classifier and explain its functionality (for correct and incorrect results) (see Fig. \ref{fig:classifiers}). After a few minutes, we showed them the data used to train the classifiers and asked them to explain how it worked, why it did not work with certain images, and to share any ideas they had for how to fix them. 

\begin{figure*}[ht]
  \centering
  \includegraphics[scale=0.7, width=\linewidth]{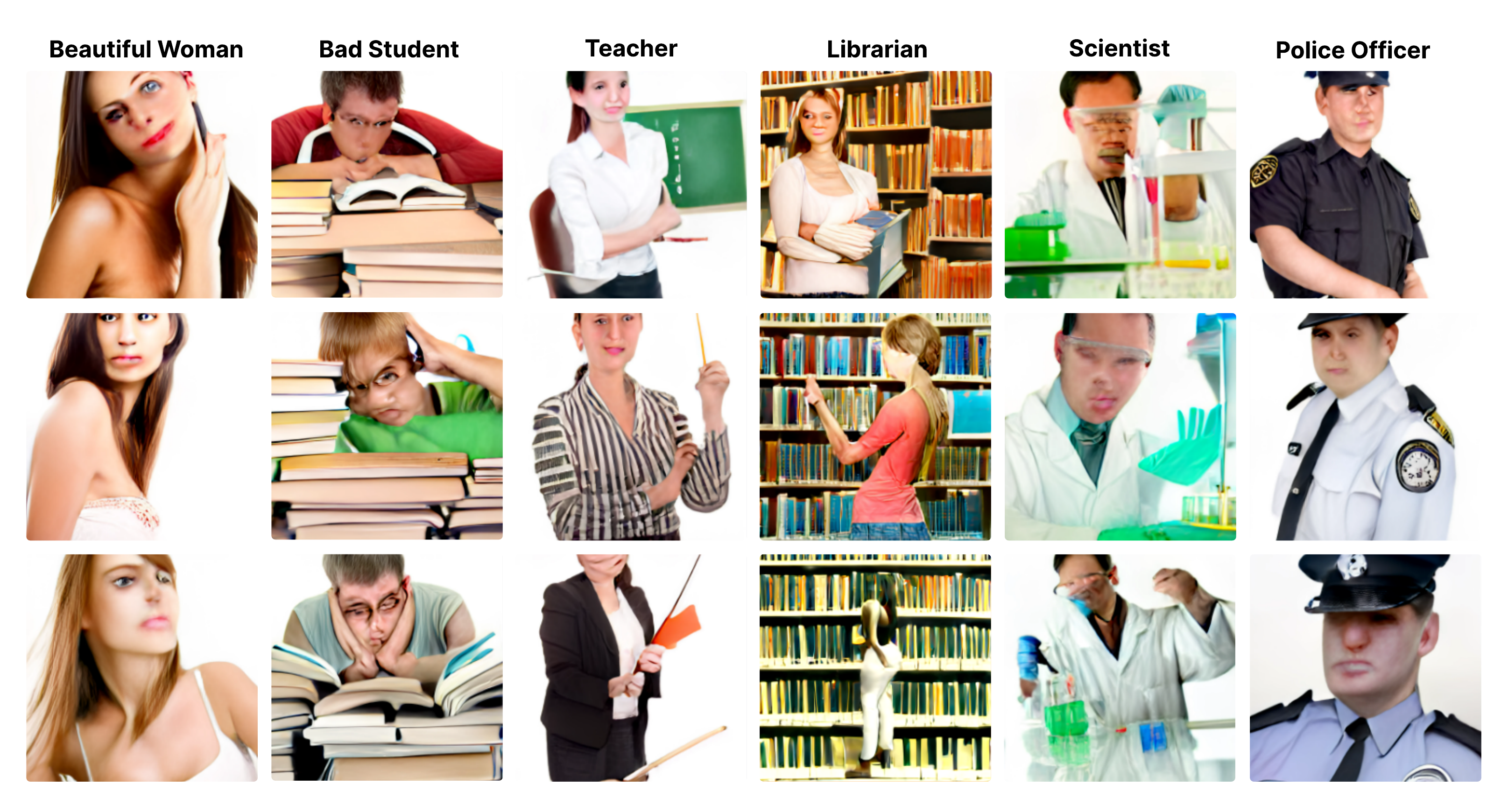}
  \caption{Images generated by DALL-E mini for the following prompts: "beautiful woman," "bad student", "teacher", "librarian", "scientist", and "police officer".}
  \Description{Images generated by DALL-E mini for the following prompts: "beautiful woman," "bad student," "teacher", "librarian", "scientist", and "police officer". All images for beautiful women are of White and skinny women with straight hair. Images of bad students are of White students surrounded by piles of books. Images of librarians include only female presenting people. Images of scientists include White males wearing coats in chemistry labs. Images of police officers are all male and White. }
  \label{fig:dalle}
\end{figure*}

For the text-to-image tasks, participants were asked to evaluate the outputs generated by DALL-E mini \cite{dayma2021dall}. This task was adapted from everyday algorithm auditing studies that have had users conduct image searches on Google to identify potentially harmful behaviors \cite{devos2022toward}, or used results from image searches and DALL-E generated images to prompt participants to reflect about algorithmic justice \cite{solyst2023potential}. In each interview, we asked participants to generate images for five topics (e.g., weddings, beautiful women, librarians, scientists) that had shown potential problematic results in previous studies and to come up with new examples that may yield problematic results (see Fig. \ref{fig:dalle}). We asked participants to share their thoughts about the results, whether they thought they were biased or discriminatory in harmful ways that might negatively impact people, and what they would do if they had the option to change or adjust the results.

Additionally, in the pre interview we asked participants if they had heard of harmful bias or discrimination in algorithmic systems or in applications used in their everyday lives and to provide any examples they were familiar with. In the post, we asked participants to tell us ``a little bit about what you learned or noticed when auditing other people’s projects" and whether auditing was helpful or not in the process of making applications and addressing potential problematic behaviors. Across both tasks and the additional question, we used follow-up prompts to elicit details from participants: “What do you think could be the causes of these issues?´´ “Can you tell me more about that?" “What makes you think so?" “Can you give me an example?" “Why did you say that?"

\subsection{Analysis}

We conducted two rounds of inductive-deductive thematic analysis \cite{braun2012thematic}. We used an automated transcription tool and then verified the transcripts for accuracy. As part of the initial analysis, two researchers inductively coded one-third of the data (consisting of three sets of matching pre/post interviews) identifying 77 emergent codes. Following this, codes were grouped to develop a codebook (9 codes and 45 subcodes); this process was also informed by previous research \cite{solyst2023potential, devos2022toward}. The codes included bias, data design, project reflection, antropomorphizing, auditing, algorithmic justice, identifying issues, prior experiences, and model design. Each code was specified through subcodes. For example, bias was broken down into subcategories, including race bias and gender bias. In a second phase of analysis, two researchers applied the coding scheme to all pre and post interviews (see codebook in Appendix \ref{appendix:codebook}). During the coding process, the researchers analyzed the data together, actively communicating with each other, discussing disagreements, and striving to reach consensus. They also had ongoing discussions with a third researcher who was knowledgeable about the data and the coding framework. We coded 1250 instances in which we observed the themes. Because this is an exploratory study with only a few participants, we focused on reaching agreement during the coding process by coding together and resolving any differences through extensive discussions rather than relying on inter-rater reliability, also in keeping with recommendations in prior work \cite{mcdonald2019reliability}.

\section{Findings}

In this section, we begin by discussing how youth identified potential algorithmic biases, then move on to their considerations of harm and justice during the interview auditing tasks. Then we discuss youth's inferences about data and model design issues, as well as their suggestions for how to address them during the interview auditing tasks. Finally, we consider youth perspectives on the benefits of peer auditing activities.

\subsection{How did youth’s identification of potential algorithmic biases and harm change? }
\subsubsection{\textbf{Identifying algorithmic biases}}

Notably, all 13 participants were able to identify potential algorithmic biases in the classifier task during the post-interview compared to 11 participants in the pre-interview. Participants identified potential biases related to body shapes, breed (in the case of animals), color, size, shape, and context/location. In the pre, for instance, Kayla argued that the pet classifier was biased against German Shepherds as it classified them as cats while other breeds of dogs (Dalmatians, Golden Retrievers, and Corgis) were correctly classified. While fewer instances in which youth discussed potential biases were identified in post (92 instances) than in pre (131 instances), in post, participants related biases to data and model design issues. We further discuss this in the next section. The only subcode in which the number of both participants and instances increased in post was for context/location bias, going from 5 participants (11 instances) to 12 participants (36 instances). Jackie Star, for example, noticed that, when testing a “tools for drawing” classifier in the post, any image that contained a hand holding the tool was classified as a pencil. She explained that the model was biased because of the context in which pencils were probably portrayed in the training data.

In the DALL-E task, overall, the number of participants that identified biases increased from 9 in pre to 12 post. Across most subcodes (age, body appearance, color, gender, context/location, race, and relevancy), the number of students that identified biases increased. Race and gender biases were the most commonly identified by participants. For racial biases, 9 participants noted these in pre and 12 in post. Fatimah, in post, argued that DALL-E was biased in favor of White people because all words related to professions (librarians, teachers, lawyers) generated images of White people. As an experiment, she tried to generate images for “thug” because “I was expecting, like, some Black dude, you know, but it was a White guy that looked like Eminem with a hood on.” This example also shows how youth’s expectations were also based on their personal biases; we discuss this below. Seven participants noticed gender biases in pre and 11 in post. In post, for example, Stephanie noted that gender in the images depended on the prompts used to generate images, with all police officers generated by DALL-E being male and all librarians being female. 

Participants (2 in pre and 5 in post) identified biases in favor of irrelevant, dated results. Lou claimed that the results were biased in favor of ``early 2000s pictures" because of the ``hairstyles and clothes" of humans generated by DALL-E as well as the ``outdated values" represented in the images. Andrés, using as an example the images generated with the prompt “beautiful woman” (see Fig. \ref{fig:dalle}), explained that the results did not represent beauty standards today because ``Lizzo, everyone calls her beautiful but none [of the generated pictures] looked like her." Fatimah and Emily also commented on how the representation of gender and sex was outdated, noting that the pictures did not include gender and sex non-conforming people and queer couples. Kayla also noticed biases in favor of older/outdated representations of humans in the pictures generated for both “good students” and “bad students,” explaining that all images included books and “chalkboards and everything” when she expected “more like actually [students] working probably like less of the books, cause most of it is digital now.” 

While completing the tasks during post, nine participants talked about trying to break the classifiers, or DALL-E, as an approach to auditing. As Iván explained, this involved “challenging it [the classifier] to see what would break it.”

When identifying potential algorithmic biases, participants sometimes reflected on their own personal biases and their perceptions of societal biases. This was particularly salient in the DALL-E task. In post 11 participants voiced being aware of their personal experiences and biases when evaluating outputs, compared with 7 participants in pre. For example, when looking at the outputs of DALL-E for teachers, Iván reflected, “from my personal experience, teaching as a very female-dominated profession.” In a similar instance, when looking at the outputs for librarians, Kayla said, “It makes sense that it will all be women? I've personally never heard of a male librarian... I really haven't. I've never seen a male librarian.” We observed 7 participants discuss how societal biases were reflected in the outputs during both pre and post. Here for example, Fatimah discussed that the outputs reflected what popular media looks like, saying that “representation in media is necessary.” In a similar instance, when looking at the outputs for gamers, Iván noted “a lot of YouTube channels it has... I feel like it's mainly run by White guy gamers.” 

\subsubsection{\textbf{Considering justice and harm}}

The number of participants who talked about justice and potential harm increased from 7 in pre to 12 in post. While completing the DALL-E tasks, youth showed diverse understandings of harm as context-dependent, being able to think about harm in terms of how they could be affected by algorithmic systems and how these could affect other people. 

A common concern was the representation of professions and how it may discourage people from pursuing certain careers. Luke explained how a lack of representation can be harmful: “For the scientists, like kids saying they want to be scientists, looking up scientists and not seeing anybody like them can kind of be like, whoa, if nobody that looks like me is a scientist, then should I really become one?” Similarly, Luke argued that beauty standards portrayed by AI-generated images could affect people’s mental health and self-esteem. 

Some participants argued that these systems could exclude people but are not necessarily harmful. In pre, Iván explained, “they're biased, they're not like making anyone look bad, but they're more like excluding people.” Notably, his perspective changed and in the post he expressed that exclusion could be potentially harmful “I feel like at this point right now, it's not harmful. But as it evolves, it will be […] if these issues aren't addressed by adding more diversity.” 
At the same time, only one participant, Richard, expressed in both pre and post that AI/ML systems could not be harmful. In pre, Richard said, “I think if you're getting harmed by an AI, I don't know, that's more of a personal problem.” In post, he explained “I don't think it can be harmful. I do think it's discriminatory. You're not gonna, like, get offended by the AI.” 

Walter and Jackie Star explained that harm depends on the context in which AI/ML systems are used. Jackie said: “Yeah, it just excludes. Like in this context, with just generating pictures. I don't know if it's really impactful.” Similarly, Walter argued that harm depends on whether “someone's using this in an actual like, like a practical use”. 

During the post interviews, participants brought up cases of algorithmic injustice they were already familiar with. Walter talked about how racial biases in image generation could also be present in how people are recognized and classified in policing systems that could be biased “towards protecting, like, White males or something like that.” Kayla also gave a similar example of how a “Black man who had never done anything wrong in his life” could be identified as a criminal in a biased facial recognition system. Lou noted that in medicine, if AI/ML systems do not recognize Black patients, it could be dangerous as people could be misdiagnosed.

\subsection{How did youth’s inferences about data and model design change?}
\subsubsection{\textbf{Making inferences}}

As participants interacted with the auditing tasks, they explained what they observed by coming up with ideas about data and model design issues that could impact the performance of the systems. In post, each participant identified an average of 12.8 possible data and model design issues, compared to an average of 6.9 issues in pre. 

In the classifier tasks, all youth identified potential model and data design-related issues in the post interview (compared to 11 participants in pre). All participants but one identified more data design issues in the post than in the pre. At the same time, the number of issues identified increased across all subcodes (i.e., model features, data composition, data diversity, data context, data sources, and class balance) except data quantity. This shows that through the workshop, in which they designed and audited applications, youth may have developed a more nuanced understanding of how data quality impacts model performance, moving beyond the popular adage that data quantity drives model performance. In the DALL-E task, the number of youth that identified data and model design issues increased from 8 in pre to 10 in post. 

In pre, participants described their understanding that models base their performance on some of the features of the data. When interacting with the pet classifier, Fatimah argued that it was important to “provide more features” to the model so that it would know what to look for and not make decisions just based on color. Other participants voiced similar ideas, talking about how the models classified images based on “key factors and traits.” Similarly, in the post, they talked about “main identifiers” and how some features “mattered more than others.”

While participants often made inferences about data diversity in general terms (11 in pre, 13 in post), in post, they referred more often to data composition (5 participants in pre, 10 in post), context (1 participant in pre, 10 in post), and sources (7 participants in pre, 8 in post). In terms of data composition, for example, Lou talked about how different camera shots influenced performance, noting that all close-ups in the sea-life classifier were classified as sharks. Jackie Star agreed, “Yeah, definitely a bias towards sharks if it was close up to a face, because that's probably all that it really is like taught on.” When looking at the data set, Richard also noted that all pictures of sharks were taken from the same angle. For the same classifier, in terms of data context, Iván noticed that all pictures of dolphins were of dolphins out of the water.  Data context was also discussed in the DALL-E task, particularly with regards to weddings, with students like Fatimah speculating that the data was probably all from the same context because “certainly with Indian weddings, there's different traditions and different ceremonies that happen for weddings, not just white dress.” The sources for the data used to train models were also discussed, with participants speculating that the data for DALL-E represented what they commonly see on certain YouTube videos (of gamers and weddings) or pictures from stock images or magazines. Jackie Star reflected that data sets are curated by humans that decide on where to source data from; it “shows more human bias than AI bias because if it's like trained off of like pictures, and that's kind of like the pictures that it's seen [...] I think it's more of like a human problem that the bots are just learning from,” she explained.

\subsubsection{\textbf{Coming up with next steps}}
Eight participants in pre and 10 participants in post came up with concrete next steps related to model and dataset design that could be taken to address the issues they identified. Next steps went from adding more data to balancing classes in pre to more nuanced ideas about data composition and augmentation in post. Walter, for example, reflected that to improve the performance of the sealife classifier, it was important to make sure each class had images composed in diverse ways. He discussed that shark images should not just include close-ups but also "zoomed out like the whole body and good lighting.” For whales, he argued that the model probably needed more pictures of whales “out of the water while jumping.” Jackie Star, in post, also voiced some ideas about data augmentation, such as rotating images or making images black and white.

\subsection{What benefits did youth find in auditing applications and having their applications audited?}
During the post interviews, youth reflected on their experiences auditing each other's applications and having their applications audited during the workshop. In particular, they valued how auditing provided them with new perspectives, gave designers ideas on how to improve projects, and helped them think about their own projects in new ways. 

Eleven participants talked about how auditing provided them with new perspectives related to model functionality and how to improve model performance and their own projects. “This is not taxes; it’s more like a game,” Richard said, describing the role of the auditor as that of someone whose goal is to identify “all the problems.” Overall, youth agreed that auditors were able to bring in new perspectives because they were unfamiliar with the projects and how these were created. Here, Iván noted that auditing involved “not just getting more diverse user input, but feedback from people that don’t think like you.” Lou explained, “you also get different standpoints because people think in so many different ways that, like, you wouldn't have thought of something and now you can incorporate that.” Luke voiced a similar idea, highlighting that auditors “may see things that [designers] have not seen.” Jerome further reflected on the collaborative nature of auditing, saying that “it's more than one perspective [...] different viewpoints come together.”

Participants also reflected that the ideas that auditors provided on how to improve projects were helpful.  Jackie Star noted that “people were like, well, you could have added more variety to this class,” giving her concrete steps on how to improve her projects. Similarly, Fatimah claimed that the feedback from auditors “helped me humble myself, helped me realize, okay, there are changes I can make, or actually my project is doing much better than I thought it would.”

Seven youth also mentioned that auditing helped them look at their own projects from different perspectives, making connections between what they saw other people do and what they were doing in their own projects. Jerome explained that after auditing, “you can turn around and improve that yourself.” Iván explained that after auditing, “I use the logic that I use in their project of challenging it to see what would break it on our project.” Lou claimed that after auditing, she was able to avoid other people’s mistakes and prevent some of the issues she observed in other people’s projects in her own project.

\section{Discussion}
In this paper, we investigated the potential benefits of positioning youth as peer auditors of AI/ML activities. Here we discuss adapting algorithm auditing methods for youth by grounding them in the rich history of CCI and the findings of our clinical interview study. 

\subsection{Peer auditing in child-computer interaction}
In the case of our study, we built on previous research that positioned youth as evaluators of their peers’ applications \cite{kafai1998children}. We observed several similarities between our work and previous work. For instance, youth benefited from cognitive distance \cite{kafai1991learning}, being able to “take perspective” \cite{ackermann2012perspective} of their own applications and those of their peers. This enabled them to provide recommendations for their peers and to apply what they saw as auditors to their own projects. Like in youth as software consultants research, playing the role of peer auditors was similar to "playing doctor" as youth identified and diagnosed issues. Yet, in our study, youth took a more adversarial approach, describing how, for some of them, the goal was to try to “break” the applications or find “all the problems”. This approach differs from the stance of expert auditors---which is about understanding systems with frequent emphasis on problematic behaviors, reflecting both the unrealistic expectations that novices may have and how their understanding of auditing may be influenced by pre-existing ideas about auditing in other fields. For instance, audits in taxation are often perceived as a threat, with people trying to avoid “being caught” by auditors \cite{advani2023dynamic, bergolo2023tax, mascagni2018lab} (such perceptions are also portrayed in popular media, e.g., Everything Everywhere All at Once). Further research is needed to better understand how non-expert auditors see their own role.

Auditing is a sociotechnical process. Our study confirms findings from previous work \cite{devos2022toward} that show that participants' interpretations about algorithmic biases are guided by their personal experiences and their understandings of societal biases. The fact that in post youth were more aware of how their personal experiences and biases influenced their perspectives of algorithmic biases suggests that auditing activities may support youth in taking perspective about both algorithmic systems and their relationship to these. This highlights the importance of thinking about auditing as sociotechnical and furthering our research not only on auditing algorithmic systems but also understanding how non-experts, including youth, audit them \cite{lam2023sociotechnical}. 

Future CCI research on youth as auditors should also build work on youth as testers and evaluators of expert-designed applications. Positioning youth as auditors of technologies designed and marketed towards them is particularly important, as they may be able to identify issues that designers and adults cannot find. At the same time, recent work conducted with adults on emergent audits, in which users evaluate systems in decentralized and distributed ways to understand their behaviors could be replicated with youth. 

\subsection{Auditing for algorithmic justice}
Our study showed that algorithm auditing tasks used in research with adults \cite{devos2022toward} cannot only be used with youth to study their perspectives towards algorithmic justice \cite{solyst2023potential}, but also be adopted in pre/post interviews to assess the potential benefits of auditing interventions. In particular, we noticed that the classifier task and the DALL-E task had unique affordances in prompting youth to think aloud about different things. The classifier task, which resembled much more closely the workshop activities (designing and auditing applications that used classifiers), prompted youth to make well-informed inferences about data and model design issues. At the same time, the DALL-E task enabled youth to make connections between what they did in the workshop and generative models. This task also prompted participants to reflect on harm by making connections to societal biases and their personal experiences. It may be more difficult to talk about issues of algorithmic justice when talking about classifying bunnies than how certain professions are represented in the outputs of a generative model.

While our findings show that even in the pre interview some participants were able to identify potential biases, it is notable that in post all participants identified potential biases. It was not surprising that some youth were able to identify potential biases in pre, as previous research shows that both adults and teenagers participating in cooperative inquiry sessions and think-aloud interviews can engage with these topics by building on their rich experiences as users of AI/ML-powered applications \cite{devos2022toward, solyst2023potential, salac2023scaffolding}. The fact that all youth identified biases and made inferences suggests the value of having youth design and audit applications. 

Like in previous work with teenagers \cite{solyst2023potential, salac2023scaffolding}, participants shared their perspectives about algorithmic justice and potential harm. After designing and auditing their peers’ applications (in post), they voiced their opinions about harm and justice more frequently. Our findings show that youth's perspectives are diverse, with some recognizing how systems could affect people in concrete ways, others arguing that harm is context-dependent or highlighting the difference between exclusion and harm, and one claiming that the burden of harm lies on the user and not algorithmic systems. These perspectives were informed by participants' positionalities (in terms of race and gender) and their lived experiences. Further research should explore how youth’s identities shape their beliefs about justice and harm. 

\subsection{Auditing and computational empowerment}

Lastly, we want to address a larger point about algorithm auditing that connects to on-going discussions about computational empowerment \cite{dindler2020computational}. Computational empowerment focuses on the construction and deconstruction of computing technologies---in our case AI/ML applications, that youth interact with. Deconstruction involves describing, evaluating and reflecting on the values and intentions embedded in sociotechnical systems and considering their possible implications \cite{dindler2020computational}. Auditing activities may be particularly well suited to support the deconstruction process. We note that all youth in the post interview were able to make inferences about data and model design. Whereas this finding is similar to those of research on youth testing their own applications \cite{tseng2023co, morales2024not} it is worth noting that the inferences were made from the outside-in, on models that participants had not designed and did not know about prior to the task. This suggests that auditing activities, beyond being helpful to identify potential harmful biases, may be productive in supporting people to understand and make sense of blackboxed AI/ML systems. The inferences made by participants show that they made connections between potential biases identified and concrete issues in the design of the models; that is, they thought about biases not as abstract but as the product of decisions made when building models in the way datasets are designed and model features and parameters are decided.

\section{Limitations}
In this paper, we used pre/post clinical interviews to investigate changes in the way youth identified bias and harm and made inferences about data and model design during auditing tasks. As such, we did not focus on the practices that youth engaged with when auditing each other's projects or the findings of the audits they conducted. Future research on peer auditing must include analyzing auditing activities microgenetically, moment-by-moment, to identify key practices and perspectives that youth may have. Such analysis could also provide useful insights into what motivates youth when auditing and what their attitudes and dispositions are towards auditing. Similarly, we did not evaluate if the issues and potential next steps proposed by youth were adequate; this should be done in future research. 

One further limitation of the findings is that we did not have a control group in which youth only designed applications. As such, it is not clear if our observations are the product of peer auditing activities, the design of applications, or both. Future studies could use the same clinical interview protocol across three treatments: one in which youth only design applications, one in which youth only audit applications, and a third one in which they design and audit applications. 

Finally, our study, like most studies related to youth and algorithmic justice, was conducted with a very small number of youth under very specific circumstances. Considering how youth identities and lived experiences may shape their beliefs about algorithmic justice, future research could intentionally sample youths with diverse experiences and backgrounds to explore how these may relate to their perspectives towards auditing. At the same time, peer auditing and youth algorithmic justice research at large should scale up and move from afterschool workshops to formal classroom settings.

\section{Conclusion}

In this paper, we introduced youth as peer auditors of AI/ML applications. Our research illustrated how youth were able not only to identify various potential biases related to gender and race but also to connect these to more complex issues of data design. Moreover, peer auditing provided youth with valuable insights for designing their own AI/ML applications. Thus, algorithm auditing expands the repertoire of roles available to children and youth in the design of computing applications in child-computer interaction research. While our study was focused on how youth conducted algorithm audits, its opportunities and limitations, and the ways they built on personal experiences, this study also points towards the possibility of including peer auditing in learning activities. Here we see a particular promise to develop algorithm auditing activities that could promote computational empowerment.  

\section{Selection and Participation of Children}
We recruited youth already enrolled in a STEM afterschool program in a city located in the Northeastern United States. Youth were invited by the organizer of the STEM program to participate via email and through paper handouts. Parents received consent forms prior to the study, which included a brief explanation of the research, and youth assented to their participation. Research protocols and data collection methods were approved by the IRB board of the University.
\begin{acks}
This analysis and writing of this paper was supported by National Science Foundation grant \#2333469. Any opinions, findings, and conclusions or recommendations expressed in this paper are those of the authors and do not necessarily reflect the views of NSF or the University of Pennsylvania. We thank Phillip Gao and Isabelle Gu for support in data collection and Deborah Fields and Lauren Vogelstein for their feedback. 
\end{acks}

\bibliographystyle{ACM-Reference-Format}
\bibliography{sample-base}


\begin{thebibliography}{68}


\ifx \showCODEN    \undefined \def \showCODEN     #1{\unskip}     \fi
\ifx \showDOI      \undefined \def \showDOI       #1{#1}\fi
\ifx \showISBNx    \undefined \def \showISBNx     #1{\unskip}     \fi
\ifx \showISBNxiii \undefined \def \showISBNxiii  #1{\unskip}     \fi
\ifx \showISSN     \undefined \def \showISSN      #1{\unskip}     \fi
\ifx \showLCCN     \undefined \def \showLCCN      #1{\unskip}     \fi
\ifx \shownote     \undefined \def \shownote      #1{#1}          \fi
\ifx \showarticletitle \undefined \def \showarticletitle #1{#1}   \fi
\ifx \showURL      \undefined \def \showURL       {\relax}        \fi
\providecommand\bibfield[2]{#2}
\providecommand\bibinfo[2]{#2}
\providecommand\natexlab[1]{#1}
\providecommand\showeprint[2][]{arXiv:#2}

\bibitem[Ackermann(2012)]%
        {ackermann2012perspective}
\bibfield{author}{\bibinfo{person}{Edith Ackermann}.} \bibinfo{year}{2012}\natexlab{}.
\newblock \showarticletitle{Perspective-taking and object construction: Two keys to learning}.
\newblock In \bibinfo{booktitle}{\emph{Constructionism in practice}}. \bibinfo{publisher}{Routledge}, \bibinfo{address}{~}, \bibinfo{pages}{25--35}.
\newblock


\bibitem[Advani et~al\mbox{.}(2023)]%
        {advani2023dynamic}
\bibfield{author}{\bibinfo{person}{Arun Advani}, \bibinfo{person}{William Elming}, {and} \bibinfo{person}{Jonathan Shaw}.} \bibinfo{year}{2023}\natexlab{}.
\newblock \showarticletitle{The dynamic effects of tax audits}.
\newblock \bibinfo{journal}{\emph{Review of Economics and Statistics}} \bibinfo{volume}{105}, \bibinfo{number}{3} (\bibinfo{year}{2023}), \bibinfo{pages}{545--561}.
\newblock


\bibitem[Ali et~al\mbox{.}(2019)]%
        {ali2019constructionism}
\bibfield{author}{\bibinfo{person}{Safinah Ali}, \bibinfo{person}{Blakeley~H Payne}, \bibinfo{person}{Randi Williams}, \bibinfo{person}{Hae~Won Park}, {and} \bibinfo{person}{Cynthia Breazeal}.} \bibinfo{year}{2019}\natexlab{}.
\newblock \showarticletitle{Constructionism, ethics, and creativity: Developing primary and middle school artificial intelligence education}. In \bibinfo{booktitle}{\emph{International workshop on education in artificial intelligence k-12 (eduai’19)}}, Vol.~\bibinfo{volume}{2}. \bibinfo{publisher}{~}, \bibinfo{address}{~}, \bibinfo{pages}{1--4}.
\newblock


\bibitem[Bandy(2021)]%
        {bandy2021problematic}
\bibfield{author}{\bibinfo{person}{Jack Bandy}.} \bibinfo{year}{2021}\natexlab{}.
\newblock \showarticletitle{Problematic machine behavior: A systematic literature review of algorithm audits}.
\newblock \bibinfo{journal}{\emph{Proceedings of the acm on human-computer interaction}} \bibinfo{volume}{5}, \bibinfo{number}{CSCW1} (\bibinfo{year}{2021}), \bibinfo{pages}{1--34}.
\newblock


\bibitem[Bergolo et~al\mbox{.}(2023)]%
        {bergolo2023tax}
\bibfield{author}{\bibinfo{person}{Marcelo Bergolo}, \bibinfo{person}{Rodrigo Ceni}, \bibinfo{person}{Guillermo Cruces}, \bibinfo{person}{Matias Giaccobasso}, {and} \bibinfo{person}{Ricardo Perez-Truglia}.} \bibinfo{year}{2023}\natexlab{}.
\newblock \showarticletitle{Tax audits as scarecrows: Evidence from a large-scale field experiment}.
\newblock \bibinfo{journal}{\emph{American Economic Journal: Economic Policy}} \bibinfo{volume}{15}, \bibinfo{number}{1} (\bibinfo{year}{2023}), \bibinfo{pages}{110--153}.
\newblock


\bibitem[Birhane(2021)]%
        {birhane2021algorithmic}
\bibfield{author}{\bibinfo{person}{Abeba Birhane}.} \bibinfo{year}{2021}\natexlab{}.
\newblock \showarticletitle{Algorithmic injustice: a relational ethics approach}.
\newblock \bibinfo{journal}{\emph{Patterns}} \bibinfo{volume}{2}, \bibinfo{number}{2} (\bibinfo{year}{2021}), \bibinfo{pages}{~}.
\newblock


\bibitem[Braun and Clarke(2012)]%
        {braun2012thematic}
\bibfield{author}{\bibinfo{person}{Virginia Braun} {and} \bibinfo{person}{Victoria Clarke}.} \bibinfo{year}{2012}\natexlab{}.
\newblock \bibinfo{booktitle}{\emph{Thematic analysis.}}
\newblock \bibinfo{publisher}{American Psychological Association}, \bibinfo{address}{~}.
\newblock


\bibitem[Buolamwini(2022)]%
        {buolamwini2022facing}
\bibfield{author}{\bibinfo{person}{Joy Buolamwini}.} \bibinfo{year}{2022}\natexlab{}.
\newblock \emph{\bibinfo{title}{Facing the Coded Gaze with Evocative Audits and Algorithmic Audits}}.
\newblock \bibinfo{thesistype}{Ph.\,D. Dissertation}. \bibinfo{school}{Massachusetts Institute of Technology}.
\newblock


\bibitem[Burhans and Dantu(2017)]%
        {burhans2017arty}
\bibfield{author}{\bibinfo{person}{Debra Burhans} {and} \bibinfo{person}{Karthik Dantu}.} \bibinfo{year}{2017}\natexlab{}.
\newblock \showarticletitle{ARTY: Fueling creativity through art, robotics and technology for youth}. In \bibinfo{booktitle}{\emph{Proceedings of the AAAI Conference on Artificial Intelligence}}, Vol.~\bibinfo{volume}{31}. \bibinfo{publisher}{~}, \bibinfo{address}{~}, \bibinfo{pages}{~}.
\newblock


\bibitem[Charisi et~al\mbox{.}(2021)]%
        {charisi2021exploring}
\bibfield{author}{\bibinfo{person}{Vicky Charisi}, \bibinfo{person}{Tomoko Imai}, \bibinfo{person}{Tiija Rinta}, \bibinfo{person}{Joy~Maliza Nakhayenze}, {and} \bibinfo{person}{Randy Gomez}.} \bibinfo{year}{2021}\natexlab{}.
\newblock \showarticletitle{Exploring the concept of fairness in everyday, imaginary and robot scenarios: a cross-cultural study with children in Japan and Uganda}. In \bibinfo{booktitle}{\emph{Proceedings of the 20th Annual ACM Interaction Design and Children Conference}}. \bibinfo{publisher}{~}, \bibinfo{address}{~}, \bibinfo{pages}{532--536}.
\newblock


\bibitem[Coenraad(2022)]%
        {coenraad2022s}
\bibfield{author}{\bibinfo{person}{Merijke Coenraad}.} \bibinfo{year}{2022}\natexlab{}.
\newblock \showarticletitle{“That’s what techquity is”: youth perceptions of technological and algorithmic bias}.
\newblock \bibinfo{journal}{\emph{Information and Learning Sciences}} \bibinfo{volume}{123}, \bibinfo{number}{7/8} (\bibinfo{year}{2022}), \bibinfo{pages}{500--525}.
\newblock


\bibitem[Dayma et~al\mbox{.}(2021)]%
        {dayma2021dall}
\bibfield{author}{\bibinfo{person}{Boris Dayma}, \bibinfo{person}{Suraj Patil}, \bibinfo{person}{Pedro Cuenca}, \bibinfo{person}{Khalid Saifullah}, \bibinfo{person}{Tanishq Abraham}, \bibinfo{person}{Ph{\'u}c Le~Khac}, \bibinfo{person}{Luke Melas}, {and} \bibinfo{person}{Ritobrata Ghosh}.} \bibinfo{year}{2021}\natexlab{}.
\newblock \showarticletitle{Dall{\textperiodcentered} e mini}.
\newblock \bibinfo{journal}{\emph{HuggingFace. com. https://huggingface. co/spaces/dallemini/dalle-mini (accessed Sep. 29, 2022)}} \bibinfo{volume}{~}, \bibinfo{number}{~} (\bibinfo{year}{2021}), \bibinfo{pages}{~}.
\newblock


\bibitem[DeVrio et~al\mbox{.}(2022)]%
        {devos2022toward}
\bibfield{author}{\bibinfo{person}{Alicia DeVrio}, \bibinfo{person}{Aditi Dhabalia}, \bibinfo{person}{Hong Shen}, \bibinfo{person}{Kenneth Holstein}, {and} \bibinfo{person}{Motahhare Eslami}.} \bibinfo{year}{2022}\natexlab{}.
\newblock \showarticletitle{Toward User-Driven Algorithm Auditing: Investigating users’ strategies for uncovering harmful algorithmic behavior}. In \bibinfo{booktitle}{\emph{Proceedings of the 2022 CHI Conference on Human Factors in Computing Systems}}. \bibinfo{publisher}{~}, \bibinfo{address}{~}, \bibinfo{pages}{1--19}.
\newblock


\bibitem[Dietz et~al\mbox{.}(2022)]%
        {dietz2022artonomous}
\bibfield{author}{\bibinfo{person}{Griffin Dietz}, \bibinfo{person}{Jennifer King~Chen}, \bibinfo{person}{Jazbo Beason}, \bibinfo{person}{Matthew Tarrow}, \bibinfo{person}{Adriana Hilliard}, {and} \bibinfo{person}{R~Benjamin Shapiro}.} \bibinfo{year}{2022}\natexlab{}.
\newblock \showarticletitle{ARtonomous: Introducing middle school students to reinforcement learning through virtual robotics}. In \bibinfo{booktitle}{\emph{Interaction Design and Children}}. \bibinfo{publisher}{~}, \bibinfo{address}{~}, \bibinfo{pages}{430--441}.
\newblock


\bibitem[Dietz et~al\mbox{.}(2020)]%
        {dietz2020giggle}
\bibfield{author}{\bibinfo{person}{Griffin Dietz}, \bibinfo{person}{Zachary Pease}, \bibinfo{person}{Brenna McNally}, {and} \bibinfo{person}{Elizabeth Foss}.} \bibinfo{year}{2020}\natexlab{}.
\newblock \showarticletitle{Giggle gauge: a self-report instrument for evaluating children's engagement with technology}. In \bibinfo{booktitle}{\emph{Proceedings of the Interaction Design and Children Conference}}. \bibinfo{publisher}{~}, \bibinfo{address}{~}, \bibinfo{pages}{614--623}.
\newblock


\bibitem[Dindler et~al\mbox{.}(2020)]%
        {dindler2020computational}
\bibfield{author}{\bibinfo{person}{Christian Dindler}, \bibinfo{person}{Rachel Smith}, {and} \bibinfo{person}{Ole~Sejer Iversen}.} \bibinfo{year}{2020}\natexlab{}.
\newblock \showarticletitle{Computational empowerment: participatory design in education}.
\newblock \bibinfo{journal}{\emph{CoDesign}} \bibinfo{volume}{16}, \bibinfo{number}{1} (\bibinfo{year}{2020}), \bibinfo{pages}{66--80}.
\newblock


\bibitem[disessa(2007)]%
        {disessa2007interactional}
\bibfield{author}{\bibinfo{person}{Andrea~A disessa}.} \bibinfo{year}{2007}\natexlab{}.
\newblock \showarticletitle{An interactional analysis of clinical interviewing}.
\newblock \bibinfo{journal}{\emph{Cognition and instruction}} \bibinfo{volume}{25}, \bibinfo{number}{4} (\bibinfo{year}{2007}), \bibinfo{pages}{523--565}.
\newblock


\bibitem[Dowthwaite et~al\mbox{.}(2020)]%
        {dowthwaite2020s}
\bibfield{author}{\bibinfo{person}{Liz Dowthwaite}, \bibinfo{person}{Helen Creswick}, \bibinfo{person}{Virginia Portillo}, \bibinfo{person}{Jun Zhao}, \bibinfo{person}{Menisha Patel}, \bibinfo{person}{Elvira~Perez Vallejos}, \bibinfo{person}{Ansgar Koene}, {and} \bibinfo{person}{Marina Jirotka}.} \bibinfo{year}{2020}\natexlab{}.
\newblock \showarticletitle{" It's your private information. it's your life." young people's views of personal data use by online technologies}. In \bibinfo{booktitle}{\emph{Proceedings of the interaction design and children conference}}. \bibinfo{publisher}{~}, \bibinfo{address}{~}, \bibinfo{pages}{121--134}.
\newblock


\bibitem[Druga and Ko(2021)]%
        {druga2021children}
\bibfield{author}{\bibinfo{person}{Stefania Druga} {and} \bibinfo{person}{Amy~J Ko}.} \bibinfo{year}{2021}\natexlab{}.
\newblock \showarticletitle{How do children’s perceptions of machine intelligence change when training and coding smart programs?}. In \bibinfo{booktitle}{\emph{Interaction design and children}}. \bibinfo{publisher}{~}, \bibinfo{address}{~}, \bibinfo{pages}{49--61}.
\newblock


\bibitem[Druin(2002)]%
        {druin2002role}
\bibfield{author}{\bibinfo{person}{Allison Druin}.} \bibinfo{year}{2002}\natexlab{}.
\newblock \showarticletitle{The role of children in the design of new technology}.
\newblock \bibinfo{journal}{\emph{Behaviour and information technology}} \bibinfo{volume}{21}, \bibinfo{number}{1} (\bibinfo{year}{2002}), \bibinfo{pages}{1--25}.
\newblock


\bibitem[Dwivedi et~al\mbox{.}(2021)]%
        {dwivedi2021exploring}
\bibfield{author}{\bibinfo{person}{Utkarsh Dwivedi}, \bibinfo{person}{Jaina Gandhi}, \bibinfo{person}{Raj Parikh}, \bibinfo{person}{Merijke Coenraad}, \bibinfo{person}{Elizabeth Bonsignore}, {and} \bibinfo{person}{Hernisa Kacorri}.} \bibinfo{year}{2021}\natexlab{}.
\newblock \showarticletitle{Exploring machine teaching with children}. In \bibinfo{booktitle}{\emph{2021 IEEE Symposium on Visual Languages and Human-Centric Computing (VL/HCC)}}. IEEE, \bibinfo{publisher}{~}, \bibinfo{address}{~}, \bibinfo{pages}{1--11}.
\newblock


\bibitem[Fails et~al\mbox{.}(2013)]%
        {fails2013methods}
\bibfield{author}{\bibinfo{person}{Jerry~Alan Fails}, \bibinfo{person}{Mona~Leigh Guha}, \bibinfo{person}{Allison Druin}, {et~al\mbox{.}}} \bibinfo{year}{2013}\natexlab{}.
\newblock \showarticletitle{Methods and techniques for involving children in the design of new technology for children}.
\newblock \bibinfo{journal}{\emph{Foundations and Trends{\textregistered} in Human--Computer Interaction}} \bibinfo{volume}{6}, \bibinfo{number}{2} (\bibinfo{year}{2013}), \bibinfo{pages}{85--166}.
\newblock


\bibitem[Fiebrink(2019)]%
        {fiebrink2019machine}
\bibfield{author}{\bibinfo{person}{Rebecca Fiebrink}.} \bibinfo{year}{2019}\natexlab{}.
\newblock \showarticletitle{Machine learning education for artists, musicians, and other creative practitioners}.
\newblock \bibinfo{journal}{\emph{ACM Transactions on Computing Education (TOCE)}} \bibinfo{volume}{19}, \bibinfo{number}{4} (\bibinfo{year}{2019}), \bibinfo{pages}{1--32}.
\newblock


\bibitem[Goldberg(1979)]%
        {goldberg1979educational}
\bibfield{author}{\bibinfo{person}{Adele Goldberg}.} \bibinfo{year}{1979}\natexlab{}.
\newblock \showarticletitle{Educational uses of a dynabook}.
\newblock \bibinfo{journal}{\emph{Computers \& Education}} \bibinfo{volume}{3}, \bibinfo{number}{4} (\bibinfo{year}{1979}), \bibinfo{pages}{247--266}.
\newblock


\bibitem[Harel(1991)]%
        {harel1991children}
\bibfield{author}{\bibinfo{person}{Idit Harel}.} \bibinfo{year}{1991}\natexlab{}.
\newblock \bibinfo{booktitle}{\emph{Children designers: Interdisciplinary constructions for learning and knowing mathematics in a computer-rich school.}}
\newblock \bibinfo{publisher}{Ablex Publishing}, \bibinfo{address}{~}.
\newblock


\bibitem[Hitron et~al\mbox{.}(2019)]%
        {hitron2019can}
\bibfield{author}{\bibinfo{person}{Tom Hitron}, \bibinfo{person}{Yoav Orlev}, \bibinfo{person}{Iddo Wald}, \bibinfo{person}{Ariel Shamir}, \bibinfo{person}{Hadas Erel}, {and} \bibinfo{person}{Oren Zuckerman}.} \bibinfo{year}{2019}\natexlab{}.
\newblock \showarticletitle{Can children understand machine learning concepts? The effect of uncovering black boxes}. In \bibinfo{booktitle}{\emph{Proceedings of the 2019 CHI conference on human factors in computing systems}}. \bibinfo{publisher}{~}, \bibinfo{address}{~}, \bibinfo{pages}{1--11}.
\newblock


\bibitem[Holbert et~al\mbox{.}(2020)]%
        {holbert2020designing}
\bibfield{author}{\bibinfo{person}{Nathan Holbert}, \bibinfo{person}{Matthew Berland}, {and} \bibinfo{person}{Yasmin~B Kafai}.} \bibinfo{year}{2020}\natexlab{}.
\newblock \bibinfo{booktitle}{\emph{Designing constructionist futures: The art, theory, and practice of learning designs}}.
\newblock \bibinfo{publisher}{MIT Press}, \bibinfo{address}{~}.
\newblock


\bibitem[Irgens et~al\mbox{.}(2022)]%
        {irgens2022characterizing}
\bibfield{author}{\bibinfo{person}{Golnaz~Arastoopour Irgens}, \bibinfo{person}{Hazel Vega}, \bibinfo{person}{Ibrahim Adisa}, {and} \bibinfo{person}{Cinamon Bailey}.} \bibinfo{year}{2022}\natexlab{}.
\newblock \showarticletitle{Characterizing children’s conceptual knowledge and computational practices in a critical machine learning educational program}.
\newblock \bibinfo{journal}{\emph{International Journal of Child-Computer Interaction}}  \bibinfo{volume}{34} (\bibinfo{year}{2022}), \bibinfo{pages}{100541}.
\newblock


\bibitem[Kafai and Harel(1991a)]%
        {kafai1991learning}
\bibfield{author}{\bibinfo{person}{Yasmin Kafai} {and} \bibinfo{person}{Idit Harel}.} \bibinfo{year}{1991}\natexlab{a}.
\newblock \showarticletitle{Learning through design and teaching: Exploring social and collaborative aspects of constructionism}.
\newblock \bibinfo{journal}{\emph{Constructionism}} \bibinfo{volume}{~}, \bibinfo{number}{~} (\bibinfo{year}{1991}), \bibinfo{pages}{85--106}.
\newblock


\bibitem[Kafai(1998)]%
        {kafai1998children}
\bibfield{author}{\bibinfo{person}{Yasmin~B Kafai}.} \bibinfo{year}{1998}\natexlab{}.
\newblock \showarticletitle{Children as designers, testers, and evaluators of educational software}.
\newblock In \bibinfo{booktitle}{\emph{The design of children's technology}}. \bibinfo{publisher}{~}, \bibinfo{address}{~}, \bibinfo{pages}{123--145}.
\newblock


\bibitem[Kafai(2012)]%
        {kafai2012minds}
\bibfield{author}{\bibinfo{person}{Yasmin~B Kafai}.} \bibinfo{year}{2012}\natexlab{}.
\newblock \bibinfo{booktitle}{\emph{Minds in play: Computer game design as a context for children's learning}}.
\newblock \bibinfo{publisher}{Routledge}, \bibinfo{address}{~}.
\newblock


\bibitem[Kafai and Harel(1991b)]%
        {kafai1991consult}
\bibfield{author}{\bibinfo{person}{Yasmin~B. Kafai} {and} \bibinfo{person}{Idit Harel}.} \bibinfo{year}{1991}\natexlab{b}.
\newblock \showarticletitle{Children's learning through consulting: When mathematical ideas, programming knowledge, instructional design, and playful discourse are intertwined.}
\newblock In \bibinfo{booktitle}{\emph{Constructionism}}. \bibinfo{publisher}{Ablex Publishing}, \bibinfo{address}{~}, \bibinfo{pages}{85--110}.
\newblock


\bibitem[Kaspersen et~al\mbox{.}(2022)]%
        {kaspersen2022high}
\bibfield{author}{\bibinfo{person}{Magnus~H{\o}holt Kaspersen}, \bibinfo{person}{Karl-Emil~Kj{\ae}r Bilstrup}, \bibinfo{person}{Maarten Van~Mechelen}, \bibinfo{person}{Arthur Hjort}, \bibinfo{person}{Niels~Olof Bouvin}, {and} \bibinfo{person}{Marianne~Graves Petersen}.} \bibinfo{year}{2022}\natexlab{}.
\newblock \showarticletitle{High school students exploring machine learning and its societal implications: Opportunities and challenges}.
\newblock \bibinfo{journal}{\emph{International Journal of Child-Computer Interaction}} \bibinfo{volume}{~}, \bibinfo{number}{~} (\bibinfo{year}{2022}), \bibinfo{pages}{~}.
\newblock


\bibitem[Kay(1996)]%
        {kay1996early}
\bibfield{author}{\bibinfo{person}{Alan~C Kay}.} \bibinfo{year}{1996}\natexlab{}.
\newblock \showarticletitle{The early history of Smalltalk}.
\newblock In \bibinfo{booktitle}{\emph{History of programming languages---II}}. \bibinfo{publisher}{~}, \bibinfo{address}{~}, \bibinfo{pages}{511--598}.
\newblock


\bibitem[Lam et~al\mbox{.}(2022)]%
        {lam2022end}
\bibfield{author}{\bibinfo{person}{Michelle~S Lam}, \bibinfo{person}{Mitchell~L Gordon}, \bibinfo{person}{Dana{\"e} Metaxa}, \bibinfo{person}{Jeffrey~T Hancock}, \bibinfo{person}{James~A Landay}, {and} \bibinfo{person}{Michael~S Bernstein}.} \bibinfo{year}{2022}\natexlab{}.
\newblock \showarticletitle{End-User Audits: A System Empowering Communities to Lead Large-Scale Investigations of Harmful Algorithmic Behavior}.
\newblock \bibinfo{journal}{\emph{Proceedings of the ACM on Human-Computer Interaction}} \bibinfo{volume}{6}, \bibinfo{number}{CSCW2} (\bibinfo{year}{2022}), \bibinfo{pages}{1--34}.
\newblock


\bibitem[Lam et~al\mbox{.}(2023)]%
        {lam2023sociotechnical}
\bibfield{author}{\bibinfo{person}{Michelle~S Lam}, \bibinfo{person}{Ayush Pandit}, \bibinfo{person}{Colin~H Kalicki}, \bibinfo{person}{Rachit Gupta}, \bibinfo{person}{Poonam Sahoo}, {and} \bibinfo{person}{Dana{\"e} Metaxa}.} \bibinfo{year}{2023}\natexlab{}.
\newblock \showarticletitle{Sociotechnical Audits: Broadening the Algorithm Auditing Lens to Investigate Targeted Advertising}.
\newblock \bibinfo{journal}{\emph{Proceedings of the ACM on Human-Computer Interaction}} \bibinfo{volume}{7}, \bibinfo{number}{CSCW2} (\bibinfo{year}{2023}), \bibinfo{pages}{1--37}.
\newblock


\bibitem[Lamichhane et~al\mbox{.}(2023)]%
        {lamichhane2023children}
\bibfield{author}{\bibinfo{person}{Dev~Raj Lamichhane}, \bibinfo{person}{Janet Read}, {and} \bibinfo{person}{Scott Mackenzie}.} \bibinfo{year}{2023}\natexlab{}.
\newblock \showarticletitle{When Children Chat with Machine Translated Text: Problems, Possibilities, Potential}. In \bibinfo{booktitle}{\emph{Proceedings of the 22nd Annual ACM Interaction Design and Children Conference}}. \bibinfo{publisher}{~}, \bibinfo{address}{~}, \bibinfo{pages}{198--209}.
\newblock


\bibitem[Lee and Fields(2017)]%
        {lee2017rubric}
\bibfield{author}{\bibinfo{person}{Victor~R Lee} {and} \bibinfo{person}{Deborah~A Fields}.} \bibinfo{year}{2017}\natexlab{}.
\newblock \showarticletitle{A rubric for describing competences in the areas of circuitry, computation, and crafting after a course using e-textiles}.
\newblock \bibinfo{journal}{\emph{The International Journal of Information and Learning Technology}} \bibinfo{volume}{34}, \bibinfo{number}{5} (\bibinfo{year}{2017}), \bibinfo{pages}{372--384}.
\newblock


\bibitem[Lehnert et~al\mbox{.}(2022)]%
        {lehnert2022child}
\bibfield{author}{\bibinfo{person}{Florence~Kristin Lehnert}, \bibinfo{person}{Jasmin Niess}, \bibinfo{person}{Carine Lallemand}, \bibinfo{person}{Panos Markopoulos}, \bibinfo{person}{Antoine Fischbach}, {and} \bibinfo{person}{Vincent Koenig}.} \bibinfo{year}{2022}\natexlab{}.
\newblock \showarticletitle{Child--Computer Interaction: From a systematic review towards an integrated understanding of interaction design methods for children}.
\newblock \bibinfo{journal}{\emph{International Journal of Child-Computer Interaction}}  \bibinfo{volume}{32} (\bibinfo{year}{2022}), \bibinfo{pages}{100398}.
\newblock


\bibitem[Long and Magerko(2020)]%
        {long2020ai}
\bibfield{author}{\bibinfo{person}{Duri Long} {and} \bibinfo{person}{Brian Magerko}.} \bibinfo{year}{2020}\natexlab{}.
\newblock \showarticletitle{What is AI literacy? Competencies and design considerations}. In \bibinfo{booktitle}{\emph{Proceedings of the 2020 CHI conference on human factors in computing systems}}. \bibinfo{publisher}{~}, \bibinfo{address}{~}, \bibinfo{pages}{1--16}.
\newblock


\bibitem[Maloney et~al\mbox{.}(2008)]%
        {maloney2008programming}
\bibfield{author}{\bibinfo{person}{John~H Maloney}, \bibinfo{person}{Kylie Peppler}, \bibinfo{person}{Yasmin Kafai}, \bibinfo{person}{Mitchel Resnick}, {and} \bibinfo{person}{Natalie Rusk}.} \bibinfo{year}{2008}\natexlab{}.
\newblock \showarticletitle{Programming by choice: urban youth learning programming with scratch}. In \bibinfo{booktitle}{\emph{Proceedings of the 39th SIGCSE technical symposium on Computer science education}}. \bibinfo{publisher}{~}, \bibinfo{address}{~}, \bibinfo{pages}{367--371}.
\newblock


\bibitem[Mascagni(2018)]%
        {mascagni2018lab}
\bibfield{author}{\bibinfo{person}{Giulia Mascagni}.} \bibinfo{year}{2018}\natexlab{}.
\newblock \showarticletitle{From the lab to the field: A review of tax experiments}.
\newblock \bibinfo{journal}{\emph{Journal of Economic Surveys}} \bibinfo{volume}{32}, \bibinfo{number}{2} (\bibinfo{year}{2018}), \bibinfo{pages}{273--301}.
\newblock


\bibitem[McDonald et~al\mbox{.}(2019)]%
        {mcdonald2019reliability}
\bibfield{author}{\bibinfo{person}{Nora McDonald}, \bibinfo{person}{Sarita Schoenebeck}, {and} \bibinfo{person}{Andrea Forte}.} \bibinfo{year}{2019}\natexlab{}.
\newblock \showarticletitle{Reliability and inter-rater reliability in qualitative research: Norms and guidelines for CSCW and HCI practice}.
\newblock \bibinfo{journal}{\emph{Proceedings of the ACM on human-computer interaction}} \bibinfo{volume}{3}, \bibinfo{number}{CSCW} (\bibinfo{year}{2019}), \bibinfo{pages}{1--23}.
\newblock


\bibitem[Metaxa et~al\mbox{.}(2021a)]%
        {metaxa2021image}
\bibfield{author}{\bibinfo{person}{Dana{\"e} Metaxa}, \bibinfo{person}{Michelle~A Gan}, \bibinfo{person}{Su Goh}, \bibinfo{person}{Jeff Hancock}, {and} \bibinfo{person}{James~A Landay}.} \bibinfo{year}{2021}\natexlab{a}.
\newblock \showarticletitle{An image of society: Gender and racial representation and impact in image search results for occupations}.
\newblock \bibinfo{journal}{\emph{Proceedings of the ACM on Human-Computer Interaction}} \bibinfo{volume}{5}, \bibinfo{number}{CSCW1} (\bibinfo{year}{2021}), \bibinfo{pages}{1--23}.
\newblock


\bibitem[Metaxa et~al\mbox{.}(2021b)]%
        {metaxa2021auditing}
\bibfield{author}{\bibinfo{person}{Dana{\"e} Metaxa}, \bibinfo{person}{Joon~Sung Park}, \bibinfo{person}{Ronald~E Robertson}, \bibinfo{person}{Karrie Karahalios}, \bibinfo{person}{Christo Wilson}, \bibinfo{person}{Jeff Hancock}, \bibinfo{person}{Christian Sandvig}, {et~al\mbox{.}}} \bibinfo{year}{2021}\natexlab{b}.
\newblock \showarticletitle{Auditing algorithms: Understanding algorithmic systems from the outside in}.
\newblock \bibinfo{journal}{\emph{Foundations and Trends{\textregistered} in Human--Computer Interaction}} \bibinfo{volume}{14}, \bibinfo{number}{4} (\bibinfo{year}{2021}), \bibinfo{pages}{272--344}.
\newblock


\bibitem[Morales-Navarro and Kafai(2023)]%
        {morales2023conceptualizing}
\bibfield{author}{\bibinfo{person}{Luis Morales-Navarro} {and} \bibinfo{person}{Yasmin~B Kafai}.} \bibinfo{year}{2023}\natexlab{}.
\newblock \showarticletitle{Conceptualizing Approaches to Critical Computing Education: Inquiry, Design, and Reimagination}.
\newblock In \bibinfo{booktitle}{\emph{Past, Present and Future of Computing Education Research: A Global Perspective}}. \bibinfo{publisher}{Springer}, \bibinfo{address}{~}, \bibinfo{pages}{521--538}.
\newblock


\bibitem[Morales-Navarro et~al\mbox{.}(2024)]%
        {morales2024not}
\bibfield{author}{\bibinfo{person}{Luis Morales-Navarro}, \bibinfo{person}{Meghan Shah}, {and} \bibinfo{person}{Yasmin~B Kafai}.} \bibinfo{year}{2024}\natexlab{}.
\newblock \showarticletitle{Not Just Training, Also Testing: High School Youths' Perspective-Taking through Peer Testing Machine Learning-Powered Applications}. In \bibinfo{booktitle}{\emph{Proceedings of the 55th ACM Technical Symposium on Computer Science Education V. 1}}. \bibinfo{publisher}{~}, \bibinfo{address}{~}, \bibinfo{pages}{~}.
\newblock


\bibitem[Robertson et~al\mbox{.}(2018)]%
        {robertson2018auditing}
\bibfield{author}{\bibinfo{person}{Ronald~E Robertson}, \bibinfo{person}{Shan Jiang}, \bibinfo{person}{Kenneth Joseph}, \bibinfo{person}{Lisa Friedland}, \bibinfo{person}{David Lazer}, {and} \bibinfo{person}{Christo Wilson}.} \bibinfo{year}{2018}\natexlab{}.
\newblock \showarticletitle{Auditing partisan audience bias within google search}.
\newblock \bibinfo{journal}{\emph{Proceedings of the ACM on Human-Computer Interaction}} \bibinfo{volume}{2}, \bibinfo{number}{CSCW} (\bibinfo{year}{2018}), \bibinfo{pages}{1--22}.
\newblock


\bibitem[Salac et~al\mbox{.}(2023a)]%
        {salac2023scaffolding}
\bibfield{author}{\bibinfo{person}{Jean Salac}, \bibinfo{person}{Rotem Landesman}, \bibinfo{person}{Stefania Druga}, {and} \bibinfo{person}{Amy~J Ko}.} \bibinfo{year}{2023}\natexlab{a}.
\newblock \showarticletitle{Scaffolding Children’s Sensemaking around Algorithmic Fairness}. In \bibinfo{booktitle}{\emph{Proceedings of the 22nd Annual ACM Interaction Design and Children Conference}}. \bibinfo{publisher}{~}, \bibinfo{address}{~}, \bibinfo{pages}{137--149}.
\newblock


\bibitem[Salac et~al\mbox{.}(2023b)]%
        {Salac2023Funds}
\bibfield{author}{\bibinfo{person}{Jean Salac}, \bibinfo{person}{Alannah Oleson}, \bibinfo{person}{Lena Armstrong}, \bibinfo{person}{Audrey Le~Meur}, {and} \bibinfo{person}{Amy~J. Ko}.} \bibinfo{year}{2023}\natexlab{b}.
\newblock \showarticletitle{Funds of Knowledge used by Adolescents of Color in Scaffolded Sensemaking around Algorithmic Fairness}. In \bibinfo{booktitle}{\emph{Proceedings of the 2023 ACM Conference on International Computing Education Research - Volume 1}} (Chicago, IL, USA) \emph{(\bibinfo{series}{ICER '23})}. \bibinfo{publisher}{Association for Computing Machinery}, \bibinfo{address}{New York, NY, USA}, \bibinfo{pages}{191–205}.
\newblock
\showISBNx{9781450399760}
\urldef\tempurl%
\url{https://doi.org/10.1145/3568813.3600110}
\showDOI{\tempurl}


\bibitem[Scaife and Rogers(1999)]%
        {scaife1999kids}
\bibfield{author}{\bibinfo{person}{Mike Scaife} {and} \bibinfo{person}{Yvonne Rogers}.} \bibinfo{year}{1999}\natexlab{}.
\newblock \showarticletitle{Kids as informants: Telling us what we didn’t know or confirming what we knew already}.
\newblock \bibinfo{journal}{\emph{The design of children’s technology}} \bibinfo{volume}{~}, \bibinfo{number}{~} (\bibinfo{year}{1999}), \bibinfo{pages}{27--50}.
\newblock


\bibitem[Shen et~al\mbox{.}(2021)]%
        {shen2021everyday}
\bibfield{author}{\bibinfo{person}{Hong Shen}, \bibinfo{person}{Alicia DeVos}, \bibinfo{person}{Motahhare Eslami}, {and} \bibinfo{person}{Kenneth Holstein}.} \bibinfo{year}{2021}\natexlab{}.
\newblock \showarticletitle{Everyday algorithm auditing: Understanding the power of everyday users in surfacing harmful algorithmic behaviors}.
\newblock \bibinfo{journal}{\emph{Proceedings of the ACM on Human-Computer Interaction}} \bibinfo{volume}{5}, \bibinfo{number}{CSCW2} (\bibinfo{year}{2021}), \bibinfo{pages}{1--29}.
\newblock


\bibitem[Sherin et~al\mbox{.}(2012)]%
        {sherin2012some}
\bibfield{author}{\bibinfo{person}{Bruce~L Sherin}, \bibinfo{person}{Moshe Krakowski}, {and} \bibinfo{person}{Victor~R Lee}.} \bibinfo{year}{2012}\natexlab{}.
\newblock \showarticletitle{Some assembly required: How scientific explanations are constructed during clinical interviews}.
\newblock \bibinfo{journal}{\emph{Journal of Research in Science Teaching}} \bibinfo{volume}{49}, \bibinfo{number}{2} (\bibinfo{year}{2012}), \bibinfo{pages}{166--198}.
\newblock


\bibitem[Solomon et~al\mbox{.}(2020)]%
        {solomon2020history}
\bibfield{author}{\bibinfo{person}{Cynthia Solomon}, \bibinfo{person}{Brian Harvey}, \bibinfo{person}{Ken Kahn}, \bibinfo{person}{Henry Lieberman}, \bibinfo{person}{Mark~L Miller}, \bibinfo{person}{Margaret Minsky}, \bibinfo{person}{Artemis Papert}, {and} \bibinfo{person}{Brian Silverman}.} \bibinfo{year}{2020}\natexlab{}.
\newblock \showarticletitle{History of logo}.
\newblock \bibinfo{journal}{\emph{Proceedings of the ACM on Programming Languages}} \bibinfo{volume}{4}, \bibinfo{number}{HOPL} (\bibinfo{year}{2020}), \bibinfo{pages}{1--66}.
\newblock


\bibitem[Solyst et~al\mbox{.}(2023a)]%
        {solyst2023would}
\bibfield{author}{\bibinfo{person}{Jaemarie Solyst}, \bibinfo{person}{Shixian Xie}, \bibinfo{person}{Ellia Yang}, \bibinfo{person}{Angela~EB Stewart}, \bibinfo{person}{Motahhare Eslami}, \bibinfo{person}{Jessica Hammer}, {and} \bibinfo{person}{Amy Ogan}.} \bibinfo{year}{2023}\natexlab{a}.
\newblock \showarticletitle{“I Would Like to Design”: Black Girls Analyzing and Ideating Fair and Accountable AI}. In \bibinfo{booktitle}{\emph{Proceedings of the 2023 CHI Conference on Human Factors in Computing Systems}}. \bibinfo{publisher}{~}, \bibinfo{address}{~}, \bibinfo{pages}{1--14}.
\newblock


\bibitem[Solyst et~al\mbox{.}(2023b)]%
        {solyst2023potential}
\bibfield{author}{\bibinfo{person}{Jaemarie Solyst}, \bibinfo{person}{Ellia Yang}, \bibinfo{person}{Shixian Xie}, \bibinfo{person}{Amy Ogan}, \bibinfo{person}{Jessica Hammer}, {and} \bibinfo{person}{Motahhare Eslami}.} \bibinfo{year}{2023}\natexlab{b}.
\newblock \showarticletitle{The Potential of Diverse Youth as Stakeholders in Identifying and Mitigating Algorithmic Bias for a Future of Fairer AI}.
\newblock \bibinfo{journal}{\emph{Proceedings of the ACM on Human-Computer Interaction}} \bibinfo{volume}{7}, \bibinfo{number}{CSCW2} (\bibinfo{year}{2023}), \bibinfo{pages}{1--27}.
\newblock


\bibitem[Tseng et~al\mbox{.}(2024)]%
        {tseng2023co}
\bibfield{author}{\bibinfo{person}{Tiffany Tseng}, \bibinfo{person}{Matt~J Davidson}, \bibinfo{person}{Luis Morales-Navarro}, \bibinfo{person}{Jennifer~King Chen}, \bibinfo{person}{Victoria Delaney}, \bibinfo{person}{Mark Leibowitz}, \bibinfo{person}{Jazbo Beason}, {and} \bibinfo{person}{R~Benjamin Shapiro}.} \bibinfo{year}{2024}\natexlab{}.
\newblock \showarticletitle{Co-ML: Collaborative Machine Learning Model Building for Developing Dataset Design Practices}.
\newblock \bibinfo{journal}{\emph{ACM Transactions on Computing Education (TOCE)}} \bibinfo{volume}{~}, \bibinfo{number}{~} (\bibinfo{year}{2024}), \bibinfo{pages}{~}.
\newblock


\bibitem[Tseng et~al\mbox{.}(2023)]%
        {tseng2023collaborative}
\bibfield{author}{\bibinfo{person}{Tiffany Tseng}, \bibinfo{person}{Jennifer King~Chen}, \bibinfo{person}{Mona Abdelrahman}, \bibinfo{person}{Mary~Beth Kery}, \bibinfo{person}{Fred Hohman}, \bibinfo{person}{Adriana Hilliard}, {and} \bibinfo{person}{R~Benjamin Shapiro}.} \bibinfo{year}{2023}\natexlab{}.
\newblock \showarticletitle{Collaborative Machine Learning Model Building with Families Using Co-ML}. In \bibinfo{booktitle}{\emph{Proceedings of the 22nd Annual ACM Interaction Design and Children Conference}}. \bibinfo{publisher}{~}, \bibinfo{address}{~}, \bibinfo{pages}{40--51}.
\newblock


\bibitem[Vakil and McKinney~de Royston(2022)]%
        {vakil2022youth}
\bibfield{author}{\bibinfo{person}{Sepehr Vakil} {and} \bibinfo{person}{Maxine McKinney~de Royston}.} \bibinfo{year}{2022}\natexlab{}.
\newblock \showarticletitle{Youth as philosophers of technology}.
\newblock \bibinfo{journal}{\emph{Mind, Culture, and Activity}} \bibinfo{volume}{29}, \bibinfo{number}{4} (\bibinfo{year}{2022}), \bibinfo{pages}{336--355}.
\newblock


\bibitem[Van~Mechelen et~al\mbox{.}(2023)]%
        {van2023emerging}
\bibfield{author}{\bibinfo{person}{Maarten Van~Mechelen}, \bibinfo{person}{Rachel~Charlotte Smith}, \bibinfo{person}{Marie-Monique Schaper}, \bibinfo{person}{Mariana Tamashiro}, \bibinfo{person}{Karl-Emil Bilstrup}, \bibinfo{person}{Mille Lunding}, \bibinfo{person}{Marianne Graves~Petersen}, {and} \bibinfo{person}{Ole Sejer~Iversen}.} \bibinfo{year}{2023}\natexlab{}.
\newblock \showarticletitle{Emerging technologies in K--12 education: A future HCI research agenda}.
\newblock \bibinfo{journal}{\emph{ACM Transactions on Computer-Human Interaction}} \bibinfo{volume}{30}, \bibinfo{number}{3} (\bibinfo{year}{2023}), \bibinfo{pages}{1--40}.
\newblock


\bibitem[Vartiainen et~al\mbox{.}(2020)]%
        {vartiainen2020learning}
\bibfield{author}{\bibinfo{person}{Henriikka Vartiainen}, \bibinfo{person}{Matti Tedre}, {and} \bibinfo{person}{Teemu Valtonen}.} \bibinfo{year}{2020}\natexlab{}.
\newblock \showarticletitle{Learning machine learning with very young children: Who is teaching whom?}
\newblock \bibinfo{journal}{\emph{International journal of child-computer interaction}}  \bibinfo{volume}{25} (\bibinfo{year}{2020}), \bibinfo{pages}{100182}.
\newblock


\bibitem[Vartiainen et~al\mbox{.}(2021)]%
        {vartiainen2021machine}
\bibfield{author}{\bibinfo{person}{Henriikka Vartiainen}, \bibinfo{person}{Tapani Toivonen}, \bibinfo{person}{Ilkka Jormanainen}, \bibinfo{person}{Juho Kahila}, \bibinfo{person}{Matti Tedre}, {and} \bibinfo{person}{Teemu Valtonen}.} \bibinfo{year}{2021}\natexlab{}.
\newblock \showarticletitle{Machine learning for middle schoolers: Learning through data-driven design}.
\newblock \bibinfo{journal}{\emph{International Journal of Child-Computer Interaction}}  \bibinfo{volume}{29} (\bibinfo{year}{2021}), \bibinfo{pages}{100281}.
\newblock


\bibitem[Voulgari et~al\mbox{.}(2021)]%
        {voulgari2021learn}
\bibfield{author}{\bibinfo{person}{Iro Voulgari}, \bibinfo{person}{Marvin Zammit}, \bibinfo{person}{Elias Stouraitis}, \bibinfo{person}{Antonios Liapis}, {and} \bibinfo{person}{Georgios Yannakakis}.} \bibinfo{year}{2021}\natexlab{}.
\newblock \showarticletitle{Learn to machine learn: designing a game based approach for teaching machine learning to primary and secondary education students}. In \bibinfo{booktitle}{\emph{Interaction design and children}}. \bibinfo{publisher}{~}, \bibinfo{address}{~}, \bibinfo{pages}{593--598}.
\newblock


\bibitem[Walker et~al\mbox{.}(2022)]%
        {walker2022liberatory}
\bibfield{author}{\bibinfo{person}{Raechel Walker}, \bibinfo{person}{Eman Sherif}, {and} \bibinfo{person}{Cynthia Breazeal}.} \bibinfo{year}{2022}\natexlab{}.
\newblock \showarticletitle{Liberatory Computing Education for African American Students}. In \bibinfo{booktitle}{\emph{2022 IEEE Conference on Research in Equitable and Sustained Participation in Engineering, Computing, and Technology (RESPECT)}}. IEEE, \bibinfo{publisher}{~}, \bibinfo{address}{~}, \bibinfo{pages}{85--89}.
\newblock


\bibitem[Yip et~al\mbox{.}(2023)]%
        {yip2023co}
\bibfield{author}{\bibinfo{person}{Jason Yip}, \bibinfo{person}{Kelly Wong}, \bibinfo{person}{Isabella Oh}, \bibinfo{person}{Farisha Sultan}, \bibinfo{person}{Wendy Roldan}, \bibinfo{person}{Kung~Jin Lee}, \bibinfo{person}{Jimi Huh}, {et~al\mbox{.}}} \bibinfo{year}{2023}\natexlab{}.
\newblock \showarticletitle{Co-design Tensions Between Parents, Children, and Researchers Regarding Mobile Health Technology Design Needs and Decisions: Case Study}.
\newblock \bibinfo{journal}{\emph{JMIR Formative Research}} \bibinfo{volume}{7}, \bibinfo{number}{1} (\bibinfo{year}{2023}), \bibinfo{pages}{e41726}.
\newblock


\bibitem[Zimmermann-Niefield et~al\mbox{.}(2020)]%
        {zimmermann2020youth}
\bibfield{author}{\bibinfo{person}{Abigail Zimmermann-Niefield}, \bibinfo{person}{Shawn Polson}, \bibinfo{person}{Celeste Moreno}, {and} \bibinfo{person}{R~Benjamin Shapiro}.} \bibinfo{year}{2020}\natexlab{}.
\newblock \showarticletitle{Youth making machine learning models for gesture-controlled interactive media}. In \bibinfo{booktitle}{\emph{Proceedings of the interaction design and children conference}}. \bibinfo{publisher}{~}, \bibinfo{address}{~}, \bibinfo{pages}{63--74}.
\newblock


\bibitem[Zimmermann-Niefield et~al\mbox{.}(2019)]%
        {zimmermann2019youth}
\bibfield{author}{\bibinfo{person}{Abigail Zimmermann-Niefield}, \bibinfo{person}{Makenna Turner}, \bibinfo{person}{Bridget Murphy}, \bibinfo{person}{Shaun~K Kane}, {and} \bibinfo{person}{R~Benjamin Shapiro}.} \bibinfo{year}{2019}\natexlab{}.
\newblock \showarticletitle{Youth learning machine learning through building models of athletic moves}. In \bibinfo{booktitle}{\emph{Proceedings of the 18th ACM international conference on interaction design and children}}. \bibinfo{publisher}{~}, \bibinfo{address}{~}, \bibinfo{pages}{121--132}.
\newblock


\bibitem[Zito et~al\mbox{.}(2021)]%
        {zito2021leveraging}
\bibfield{author}{\bibinfo{person}{Lauren Zito}, \bibinfo{person}{Jennifer~L Cross}, \bibinfo{person}{Bambi Brewer}, \bibinfo{person}{Samantha Speer}, \bibinfo{person}{Michael Tasota}, \bibinfo{person}{Emily Hamner}, \bibinfo{person}{Molly Johnson}, \bibinfo{person}{Tom Lauwers}, {and} \bibinfo{person}{Illah Nourbakhsh}.} \bibinfo{year}{2021}\natexlab{}.
\newblock \showarticletitle{Leveraging tangible interfaces in primary school math: Pilot testing of the Owlet math program}.
\newblock \bibinfo{journal}{\emph{International Journal of Child-Computer Interaction}}  \bibinfo{volume}{27} (\bibinfo{year}{2021}), \bibinfo{pages}{100222}.
\newblock


\end{thebibliography}

\appendix

\section{Codebook}
\label{appendix:codebook}
\footnotesize

\textbf{anthropomorphizing}
\begin{itemize}
  \item Definition: Attaching human characteristics to the model.
  \item Example: “It's thinking this is a shark” Richard
\end{itemize}

\textbf{auditing.breakIt}
\begin{itemize}
  \item Definition: Indicating that auditing involves finding moments when the project breaks.
  \item Example: “Challenging it [the project] to see what would break it ” Iván
\end{itemize}

\textbf{auditing.newPerspectives}
\begin{itemize}
  \item Definition: Voicing that auditors provide new perspectives on how the project works.
  \item Example: “it is helpful to get feedback and other perspectives from other people. Who may see things that we have not seen.” Luke
\end{itemize}

\textbf{auditing.nextSteps}
\begin{itemize}
  \item Definition: Coming up with next steps to improve model performance.
  \item Example: “Add more variety in the data. Like these are all the same kind of rabbit.” Kayla
\end{itemize}

\textbf{auditing.noticePatterns}
\begin{itemize}
  \item Definition: Identifying patterns in the outputs to build explanations for model behaviors.
  \item Example: “which I feel like that's kind of customary for DALL-E because I feel like that the images it's been fed with are more probably just white people instead of diverse like images.” Iván
\end{itemize}

\textbf{auditing.applyingKnowledge}
\begin{itemize}
  \item Definition: Indicating that insights gained while auditing peers’ projects can be helpful to improve their own projects.
  \item Example: “I felt like just going in, and being able to see other people's projects and see what they can improve on. It's like, what can I improve on it is like, your telling somebody else, what they can improve on and you can turn around and improve that yourself.” Jerome
\end{itemize}

\textbf{audtiting.placeResponsibility}
\begin{itemize}
  \item Definition: Referring to something or someone as responsible for the outputs.
  \item Example: “I think it shows more of like human bias than like, AI bias maybe that that's like what like the like, because if it's like trained off of like pictures, like you were showing, and that's kind of like the pictures that it's been like that it's seen being put up on the internet.” Jackie Star
\end{itemize}

\textbf{bias.age}
\begin{itemize}
  \item Definition: Identifying potential age related biases in model outputs.
  \item Example: “it's all like let's say young woman like for like the other ones there's like more young people but like they should I feel like they should I add like more older people are like mid-age these are these people look really young” Walter
\end{itemize}

\textbf{bias.appearance.body}
\begin{itemize}
  \item Definition: Identifying potential body appearance related biases in model outputs.
  \item Example: “it's all like let's say young woman like for like the other ones there's like more young people but like they should I feel like they should I add like more older people are like mid-age these are these people look really young” Walter
\end{itemize}

\textbf{bias.appearance.body}
\begin{itemize}
  \item Definition: Identifying potential fashion related biases in model outputs.
  \item Example: “like the general like, in the lab, mixing stuff together, lab coats, goggles gloves. Again, is mainly just just white people.” Jerome
\end{itemize}

\textbf{bias.breed}
\begin{itemize}
  \item Definition: Identifying potential breed related biases in model outputs.
  \item Example: “Yeah like the Corgi you can see is all... the Corgi is [identified as] a dog and the German Shepherd as the cat. Why would it think that?” Kayla
\end{itemize}

\textbf{bias.color}
\begin{itemize}
  \item Definition: Identifying potential breed related biases in model outputs.
  \item Example: “Yeah like the Corgi you can see is all... the Corgi is [identified as] a dog and the German Shepherd as the cat. Why would it think that?” Kayla
\end{itemize}

\textbf{bias.gender}
\begin{itemize}
  \item Definition: Identifying potential gender related biases in model outputs.
  \item Example: “The models were all white girls with straight hair..” Jackie Star
\end{itemize}

\textbf{bias.location/context}
\begin{itemize}
  \item Definition: Identifying potential location/context related biases in model outputs.
  \item Example: “All the dolphins are jumping on the water none of this whales are so it's creating a bias towards the dolphins because they're all technically in the same act that having the same like actions and as dolphins.” Fatimah
\end{itemize}

\textbf{bias.position}
\begin{itemize}
  \item Definition: Identifying potential position related biases in model outputs.
  \item Example: “These cats are mainly perched up. Yeah, I feel like in the same position.” Emily
\end{itemize}

\textbf{bias.race}
\begin{itemize}
  \item Definition: Identifying potential race related biases in model outputs.
  \item Example: “For the doctor that was always white male, or any other professions like lawyer teacher well teachers mean female or like librarians female but it was majority white people and no black or any race.” Fatimah
\end{itemize}

\textbf{bias.relevancy}
\begin{itemize}
  \item Definition: Identifying potential relevancy related biases in model outputs.
  \item Example: “This was probably how what a good student was like defined as back in the day.” Stephanie
\end{itemize}

\textbf{bias.shape}
\begin{itemize}
  \item Definition: Identifying potential shape related biases in model outputs.
  \item Example: “Because I had just like a prominent shape like yeah, shape or triangles and cones. It's easy to see compared to the other two, which can be kind of a bit similar. It made the [blackberry] look like strawberries.” Kayla
\end{itemize}

\textbf{bias.size}
\begin{itemize}
  \item Definition: Identifying potential size related biases in model outputs.
  \item Example: “this thing just thinks everything that has big ears as a cat.” Richard
\end{itemize}

\textbf{bias.socioeconomic}
\begin{itemize}
  \item Definition: Identifying potential biases related to socioeconomic status in model outputs.
  \item Example: “They're all like white again, like probably middle class... upper middle class white people... mainly white men or boys I guess. ” Kayla
\end{itemize}

\textbf{dataDesign.classBalance}
\begin{itemize}
  \item Definition: Inferring data composition related issues in the design of training datasets.  
  \item Example: “I would add more pictures of like sharks. Like zoomed out like the whole body color, good lighting and the whale add more like out of water where it is jumping maybe like the dolphins nothing because it seems to get like a dolphins pretty accurately.” Walter
\end{itemize}

\textbf{dataDesign.context}
\begin{itemize}
  \item Definition: Inferring data context related issues in the design of training datasets.  
  \item Example: “There may be like hands involved or like, like a background, like a lot of greenery so it knows what's the difference... because this one's just covered, the background is just white. So there's nothing, nothing in the background.” Walter
\end{itemize}

\textbf{dataDesign.diversity}
\begin{itemize}
  \item Definition: Inferring data diversity related issues in the design of training datasets.   
  \item Example: “More representation definitely just different environments different people different skin colors, races genders and even for this maybe like same sex couples that could be something” Fatimah
\end{itemize}

\textbf{dataDesign.edgeCase}
\begin{itemize}
  \item Definition: Identifying potential edge cases.   
  \item Example: “I don't think it's really used to like pencils or markers that are like black and white because all the other examples are brightly colored yeah.” Jackie Star
\end{itemize}

\textbf{dataDesign.lighting}
\begin{itemize}
  \item Definition: Inferring lighting related issues in the design of training datasets.   
  \item Example: “Like better lighting for this one, because they're all like underwater.” Kayla
\end{itemize}

\textbf{dataDesign.quantity}
\begin{itemize}
  \item Definition: Inferring data quantity related issues in the design of training datasets.   
  \item Example: “We would add more hands to holding new markers, we will do some darker markers. There's only like two darker markers.” Emily
\end{itemize}

\textbf{dataDesign.source}
\begin{itemize}
  \item Definition: Inferring issues related to the sources of data used in training datasets.  
  \item Example: “Maybe based off of like stock images, like that's what it was trained on.” Kayla
\end{itemize}

\textbf{ID.failure.explanation}
\begin{itemize}
  \item Definition: Identifying cases in which the model did not perform as expected and providing an explanation for such performance. 
  \item Example: “Yeah, it's so they go based off of the sort of since the cat is white. It goes to the bunny since all of the pictures of the bunny are white.” Luke
\end{itemize}

\textbf{ID.failure.noExplanation}
\begin{itemize}
  \item Definition: Identifying cases in which the model did not perform as expected without providing an explanation for such performance. 
  \item Example: “That's not a shark.” Luke 
\end{itemize}

\textbf{ID.success}
\begin{itemize}
  \item Definition: Identifying cases in which the model performs as expected.
  \item Example: “They all have like the big ears. I think that's why they got these two right because they have big dog ears.” Kayla
\end{itemize}

\textbf{justice.exclusion}
\begin{itemize}
  \item Definition: Arguing that system outputs exclude some people. 
  \item Example: “Because people can look at this, these pictures of doctors and rich people and see that they look nothing like them? And then feel discouraged? As if Oh, that's not an opportunity that I can convey?” Luke
\end{itemize}

\textbf{justice.harmful}
\begin{itemize}
  \item Definition: Arguing that system outputs produce harm. 
  \item Example: “Mainly the librarian one is harmful because it shows a bunch of women... it's not like men can be librarians, "don't do that... it's not a masculine job, you shouldn't have that job if you're a man." I guess it's kind of saying.” Kayla
\end{itemize}

\textbf{justice.notHarmful}
\begin{itemize}
  \item Definition: Arguing that system outputs do not produce harm. 
  \item Example: “I think if you're getting harmed by an AI, I don't know.  That's more of a  personal problem.” Richard
\end{itemize}

\textbf{justice.potentiallyHarmful}
\begin{itemize}
  \item Definition: Arguing that system outputs could be potentially harmful. 
  \item Example: “I feel like at this point right now, it's not harmful but like, as it evolves, it will be is like beautiful. I feel like no one's actually going to compare themselves to these awful images, and be like, Wow, if I don't if I don't look like this person, my God and of the world, but I feel like as it evolves, it will be more harmful if these issues aren't addressed by adding more diversity.“ Iván
\end{itemize}

\textbf{modelDesign.features}
\begin{itemize}
  \item Definition: Inferring parameters or features used by the model. 
  \item Example: “I don't think it's taking in color, it doesn't care if the strawberry is white or whatever the color of strawberries it's more looking at the texture.” Richard
\end{itemize}

\textbf{priorExperiences.personalBias}
\begin{itemize}
  \item Definition: Considering personal biases in evaluating system behaviors.
  \item Example: “It makes sense that it will all be woman. I've personally never heard of a male librarian... I really haven't.” Kayla
\end{itemize}

\textbf{priorExperiences.societalBias}
\begin{itemize}
  \item Definition: Considering societal biases in evaluating system behaviors.
  \item Example: “What about like cook because I feel like that could go in either direction. Yeah, because it could be stereotyped with women being in the kitchen…” Iván
\end{itemize}

\textbf{programmed}
\begin{itemize}
  \item Definition: Indicating that the machine has been programmed to work in a certain way.
  \item Example: “mean, I've seen... I saw like this one thing, I don't remember where it was, it was like, it asked the chat GPT to do a prompt for a smart scientist. And it just put out that the scientists had to be white. And I found that was kind of interesting, because it was that was the biases that it was programmed with, when that's not true.” Iván
\end{itemize}

\end{document}